\documentclass[a4paper, 12 pt]{article}
\usepackage[T1]{fontenc}
\usepackage[utf8]{inputenc}
\usepackage[english]{babel}
\usepackage{amsmath}
\usepackage{amssymb}
\usepackage{braket}
\usepackage{mathtools}
\usepackage{dsfont}
\usepackage{subcaption}
\usepackage{array}
\usepackage{booktabs}
\usepackage{yfonts}
\usepackage{setspace}
\usepackage{verbatim}
\usepackage{geometry}
\usepackage[usenames,dvipsnames]{color}
\usepackage{hyperref}
\hypersetup{
  colorlinks,
  citecolor=Blue,
  linkcolor=Blue,
  urlcolor=Blue}

\numberwithin{equation}{section}

\def\be{\begin{equation}}
\def\ee{\end{equation}}
\def\rme{{\rm e}}
\newcommand{\nn}{\nonumber}
\newcommand{\diff}{\mathrm{d}}
\newcommand{\ii}{i}

\newcommand{\vev}[1]{\hat{#1}}

\begin{document}
\pagestyle{empty}

\begin{center}

$\,$
\vskip 1.5cm

{\LARGE{\bf Gravitational index, black hole saddle\\[3mm] degeneracy, and one-form symmetry
}}

\vskip 1cm

{
{\bf Davide Cassani,${}^{\textrm 1}$
Gianmarco Esposto${}^{\textrm{1,2}}$ }
}

\vskip 1cm

\end{center}

\renewcommand{\thefootnote}{\arabic{footnote}}

\begin{center}
$^{\textrm 1}$ {\it INFN, Sezione di Padova, Via Marzolo 8, 35131 Padova, Italy},\\ [2mm] 
$^{\textrm 2}${\it Dipartimento di Fisica e Astronomia ``Galileo Galilei'', Universit\`a di Padova,\\Via Marzolo 8, 35131 Padova, Italy}\\[2mm]

\vskip 3cm

 {\bf Abstract} 
\end{center}

{\noindent It is known that the saddle describing the contribution of supersymmetric black holes to the index of four-dimensional superconformal field theories is degenerate. This degeneracy is a consequence of the one-form symmetry of the theory, and can also be seen as a specific logarithmic correction to the saddle-point action. We discuss the gravitational realization of the one-form symmetry and of the index-saddle degeneracy in different holographic setups. In order to illustrate the spontaneous breaking of the one-form symmetry at infinite volume, we employ a gravitational realization of the Cardy-like limit of small chemical potentials where the black hole decompactifies into a black brane.
}

\newpage
\setcounter{page}{1}
\pagestyle{plain}

\tableofcontents

\newpage

\section{Introduction}

Higher-form symmetries~\cite{Gaiotto:2014kfa}, whose charged objects are extended operators such as lines and surfaces, have been reshaping our understanding of topological and other global features of quantum field theories in recent years. A well-known example is the discrete one-form symmetry acting on Wilson loops in four-dimensional gauge theories, which generalizes the notion of center symmetry and serves as a diagnostic for confinement. We denote by $G$ this discrete one-form symmetry and by $|G|$ its order.
At finite temperature, Wilson loops that wind around the Euclidean time circle---also known as Polyakov loops---measure the free energy required to insert a single static quark. As such, they provide an order parameter for the confinement-deconfinement transition.
In the confining phase, the free energy of an isolated quark is infinite, and the Polyakov loop therefore has a vanishing expectation value; correspondingly, the one-form symmetry $G$ is preserved. In the deconfined phase, by contrast, the free energy is finite, leading to a nonzero expectation value for the Polyakov loop. This signals the spontaneous breaking of the one-form symmetry. It follows that in the deconfined phase, the theory on $S^1\times \mathbb{R}^3$, where $S^1$ is the thermal circle, has $|G|$ degenerate vacua. When the theory is considered at finite volume instead, namely when the spatial $\mathbb{R}^3$ is replaced by a compact manifold such as $S^3$, we should sum over the degenerate vacua in the canonical partition function ${\rm Tr}\,\rme^{-\beta E}$. The sum restores the symmetry and gives $|G|$ times the contribution of one vacuum. 

In holography, the confinement-deconfinement phase transition of a large-$N$ conformal field theory is realized by the Hawking--Page transition~\cite{Hawking:1982dh} between thermal Anti de Sitter (AdS) space and a Euclidean AdS black hole satisfying the same boundary conditions \cite{Witten:1998qj,Witten:1998zw,Aharony:1998qu}. The bulk counterparts of Polyakov loops are string worldsheets 
  wrapping the Euclidean time circle at the boundary and extending in the bulk. The Polyakov loop expectation value at large $N$ is given by the (renormalized) Euclidean string action $I_{\rm string}$ via the formula $\langle P \rangle\, \sim\, \rme^{-I_{\rm string}}$, with the different topology of Euclidean AdS and the black hole leading to different results. In empty AdS, 
the Euclidean time circle is non-contractible, implying that $I_{\rm string}$ diverges and thus that the expectation value vanishes. By contrast, in the finite-temperature black hole topology the Euclidean time circle is contractible, and $I_{\rm string}$ is finite.

In order to describe a global discrete one-form symmetry at the boundary, we must consider a discrete two-form gauge theory in the bulk. This can be modeled as a theory of a pair of flat two-form gauge fields, $b_2$, $c_2$, in five dimensions \cite{Aharony:1998qu,Gross:1998gk,Hofman:2017vwr,Bah:2020uev,Brennan:2023mmt,Bhardwaj:2023kri}. We assume that $b_2$ is the field gauging the electric one-form symmetry transformations, under which Polyakov loops are charged. (Namely, the string couples to $b_2$, implying that the action  $I_{\rm string}$ shifts under large gauge transformations of the latter, realizing the action of the one-form symmetry on the loop).  In the bulk action, these flat two-forms can only affect Chern–Simons terms, and one can focus on the BF coupling
\begin{align}\label{csterm}
S_{\text{CS}} \,=\, \frac{|G|}{2\pi} \int_{M_5} b_2 \wedge \diff c_2\,.
\end{align}
This plays a central role in the argument proving that  the symmetry is discrete and has order $|G|$ 
\cite{Witten:1998xy,Gross:1998gk,Aharony:1998qu}.
Precisely terms of this form arise when compactifying ten- or eleven-dimensional supergravity to five dimensions, with the normalization being fixed by integration over the internal manifold. Also, since
 \eqref{csterm} comes with just one derivative, it makes sense to ignore the kinetic terms and all higher-derivative terms if we ask questions about the very low-energy regime.

For $\mathcal{N}=4$ SYM theory with gauge group ${\rm SU}(N)$, for instance, the one-form symmetry  is just the $\mathbb{Z}_N$ center symmetry (we will sometimes use the notation $\mathbb{Z}_N^{(1)}$), hence $|G|=N$ in this case. The realization of this symmetry in the dual type IIB string theory on $S^5$ was discussed  in the early days of the AdS/CFT correspondence in~\cite{Aharony:1998qu}. In this case, the level $|G|$ of the Chern--Simons term \eqref{csterm} arises from the RR five-form flux over $S^5$. The field gauging the one-form symmetry in the bulk is the NSNS two-form $B$ and the holographic Polyakov loops are realized by fundamental strings. 

As in the evaluation of the string action, the topology of the Euclidean black hole plays a crucial role in realizing the $|G|$ distinct configurations corresponding to the action of the one-form symmetry. The contractible Euclidean time circle makes it possible to turn on flat $b_2$ field configurations with a non-vanishing integral $\int_{\Sigma_2} b_2$ over the cigar-like two-cycle $\Sigma_2$ spanned by the radial and Euclidean time directions. The one-form symmetry acts as a shift $\int_{\Sigma_2} b_2  \to \int_{\Sigma_2} b_2 + \frac{2\pi k}{ |G|}\,,$ $ k =0,1,\ldots,|G|-1 $. The $|G|$ configurations are distinguished by the value of $\int_{\Sigma_2} b_2$ and are degenerate, namely they carry the same on-shell action. This becomes clear upon quantizing the  fluctuations of $c_2$ around a given background~\cite{Aharony:1998qu}.

When considering the gravitational path integral with boundary conditions leading to a canonical partition function~\cite{Gibbons:1976ue}, the arguments above lead to conclude that the semiclassical saddle-point corresponding to the Euclidean AdS$_5$ black hole  must be degenerate, with degeneracy equal to $|G|$. 
 By the same reasoning as in field theory, the black hole saddle thus gives $|G|$ times the same contribution to the canonical partition function $Z$, that is
 \be
 Z \,=\, |G|\, \rme^{-I_{\rm BH}} + \ldots\,,
 \ee
 where $I_{\rm BH}$ is the black hole Euclidean action and the dots denote other corrections, including contributions from other saddles. 
 This $|G|$-fold degeneracy can also be seen as a logarithmic correction to the black hole saddle-point action, 
 \be\label{eq:log_correction}
 Z \,=\, \rme^{-I_{\rm BH} + \log |G|} + \ldots \,.
 \ee
This is a universal contribution to the logarithmic correction to the AdS$_5$ black hole action and therefore---after a change of statistical ensemble---to the black hole entropy.

Building on the general framework outlined above, in this paper we investigate the role of one-form symmetry in relation to black hole saddles of superconformal indices and the counting of supersymmetric black hole microstates, focusing on its gravitational description. Our discussion will include the case of type IIB on $S^5$ and the dual $\mathcal{N}=4$ SYM theory, but will hold much more generally.

Over the past few years, starting with \cite{Cabo-Bizet:2018ehj,Choi:2018hmj,Benini:2018ywd}, it has become clear that the superconformal index counting BPS states in four-dimensional superconformal field theories (SCFTs)~\cite{Kinney:2005ej,Romelsberger:2005eg} contains a  large-$N$ contribution that precisely reproduces the on-shell action of supersymmetric black holes. Upon performing a Legendre transform~\cite{Hosseini:2017mds}, this contribution accounts for the corresponding black hole entropy. 
This can be seen by different methods, including the analysis of large-$N$ saddles of the superconformal index and the study of a Cardy-like regime of small chemical potentials, 
 which is valid at finite~$N$. 

As we will review, the SCFT analysis further reveals that the saddle-point action associated with the black hole contribution to the superconformal index comes with a logarithmic  correction exactly equal to the $\log|G|$ term appearing in~\eqref{eq:log_correction}, which in  many  examples (beyond ${\rm SU}(N)$ $\mathcal{N}=4$ SYM) is proportional to $\log N$.
 This means that the entire $\log N$ correction to the saddle-point action is accounted for by the degeneracy of the black hole index saddle that follows from the one-form symmetry $G$. There is no net contribution to this term arising from small quantum fluctuations around the saddle.\footnote{Note that the $\log N$ correction to the black hole index saddle contributes to but does not coincide with the $\log N$ correction to the black hole entropy. Indeed, when performing the Laplace transform taking from the grand-canonical to the microcanonical ensemble one  picks up an additional $\log N$ contribution \cite{David:2021qaa}. The overall logarithmic correction to the black hole entropy obtained in this way has been matched with the result of an independent computation based on a CFT$_2$ Cardy formula in the context of the Kerr/CFT correspondence~\cite{David:2021qaa}.}
In the following, we will explain the gravitational counterpart of these statements and discuss the bulk origin of the  logarithmic term $\log|G|$.\footnote{We will not attempt here---although it would certainly be interesting---to verify through a direct bulk computation that no $\log N$ contribution arises from the one-loop determinant of supergravity field fluctuations around the asymptotically AdS$_5$ black hole saddle. We emphasize that this feature is peculiar to the setup we are considering. In different setups, the one-loop determinants of supergravity field fluctuations are known to give non-trivial logarithmic corrections both to the gravitational action and black hole entropy, see e.g.~\cite{Banerjee:2011jp,Sen:2012cj,Bhattacharyya:2012ye,Liu:2017vbl,Bobev:2023dwx}.}

The gravitational counterpart of the superconformal index can be defined by imposing appropriate boundary conditions on the supergravity path integral, such that only supersymmetric asymptotically AdS$_5$ field configurations contribute. This quantity is known as the {\it gravitational index} and has attracted considerable attention recently; see e.g.~\cite{Cassani:2025sim} for a discussion.

While, as mentioned above, the origin of black hole saddle degeneracy is largely topological from a bulk perspective, the supersymmetric setting in which the gravitational index is defined introduces several distinctive features. In particular, supersymmetric AdS$_5$ black holes carry both angular momentum and electric charge, while the analysis of~\cite{Aharony:1998qu} just considered the Schwarzschild-AdS$_5$ black hole. Moreover, the black hole on-shell action is calculated by deforming the supersymmetric solution away from extremality~\cite{Cabo-Bizet:2018ehj}. This provides regular saddles to the gravitational index and---crucially for our purposes---restores a capped disc topology in the geometry, unlike the infinitely long throat characteristic of the extremal case. At the same time, however, it necessitates considering complex field configurations on this topology. Our aim is to explain how these features fit into the broader picture.

We will discuss in detail the case of type IIB supergravity
on 5d Sasaki–Einstein manifolds (which includes $S^5$). However, we will also consider other setups allowing for supersymmetric AdS$_5$ solutions, 
both in eleven-dimensional and in massive IIA supergravity.  The supersymmetric black hole we consider is the solution to minimal five-dimensional gauged supergravity found in~\cite{Chong:2005da}, which uplifts to all such higher-dimensional setups.

We also study the limit where the boundary $S^3$ has infinite volume, revisiting the analysis of~\cite{Aharony:1998qu}. This analysis is based on a reduction on the disc to three dimensions, and retains some of the two-derivative terms in the bulk action in addition to the Chern--Simons term.
In order to discuss the infinite-volume setup, we introduce a gravitational counterpart of the Cardy-like limit which is used in the SCFT analysis. This is obtained by taking the horizon radius $r_+$ of the supersymmetric non-extremal black hole saddle to be very large, together with an appropriate rescaling of the coordinates. 
As a result, the $S^3$ that appears both at the horizon and at the boundary is effectively decompactified to $\mathbb{R}^3$, yielding a supersymmetric black brane geometry. Importantly, this scaling solution is still a non-extremal configuration, where the Euclidean time circle has finite length at the boundary and collapses smoothly at the horizon. This makes it possible to turn on flat $b_2$ field configurations with finite integral over the disc.

We then perform the analysis of quantum fluctuations around the scaling solution. In order to retain the relevant topological BF-like coupling \eqref{csterm} in five dimensions, we adopt a consistent truncation of type IIB supergravity on Sasaki--Einstein manifolds that keeps a richer (but finite) set of fields than those appearing in minimal gauged supergravity. In particular, the truncation retains the five-dimensional two-form fields $b_2$ and $c_2$, which descend from the ten-dimensional NSNS and RR two-forms, respectively. These fields are precisely the degrees of freedom needed to realize the topological sector responsible for the saddle degeneracy. This setup, including the two-derivative terms, allows us to determine the number of vacua by minimizing the resulting effective potential. We find that the gravitational Cardy-like limit removes the effects of rotation and the vacuum structure is insensitive to the black-hole charge parameter. We thus confirm that the resulting degeneracy is indeed topological and independent of the detailed local geometry.

The rest of the paper is structured as follows. In section~\ref{sec:index_saddles_and_1form} we review the field theory side of the story, discussing the origin of the logarithmic term from the one-form symmetry. In section~\ref{sec3} we present our gravitational counterpart of the Cardy-like limit, where the boundary $S^3$ decompactifies to $\mathbb{R}^3$, and we apply it to the supersymmetric non-extremal black hole saddle of minimal gauged supergravity. Then, in section~\ref{sec:GravityDegeneracy} we illustrate the gravitational origin of black hole saddle degeneracy, both at finite and infinite volume. We conclude in section~\ref{sec:conlcusions} with an outlook, discussing in particular the effect of gauging the electric one-form symmetry at the boundary. The appendix contains some more technical aspects of our analysis.

\section{Degeneracy of index saddles and one-form symmetry}\label{sec:index_saddles_and_1form}

In this section we illustrate the role of the one-form symmetry in the SCFT index. We discuss how the saddle degeneracy arises in two regimes that have played a central role in reproducing the entropy of BPS AdS$_5$ black holes, namely the Cardy-like limit and the large-$N$ limit.

\subsection{The superconformal index}

 We consider an $\mathcal{N}=1$
SCFT on the spatial manifold $S^3$. The superconformal index~\cite{Kinney:2005ej,Romelsberger:2005eg} is a
refined Witten index, defined as
\begin{equation}
\mathcal{I}(\omega_1 + 2\pi i n_0, \omega_2)
\,=\, \mathrm{Tr}\, (-1)^F \rme^{-\beta\{\mathcal{Q},\mathcal{Q}^\dagger\}
-(\omega_1+2\pi i n_0)\left(J_1+\frac{1}{2}R\right)
-\omega_2\left(J_2+\frac{1}{2}R\right)}\,,
\label{eq:indexdef}
\end{equation}
where the trace is taken over the Hilbert space of the theory on $S^3$, $F$ is the fermion number, $R$ is the R-charge, $J_1,J_2$ are angular momenta corresponding to the Cartan elements of the $SO(4)$ isometry of round $S^3$, and $\mathcal{Q}$ is a complex supercharge commuting both with $J_1+\tfrac{1}{2}R$ and $J_2+\tfrac{1}{2}R$. The complex variables $\omega_1+2\pi i n_0$, $\omega_2$ are chemical potentials for these two charge combinations. As in \cite{Cabo-Bizet:2019osg,Cabo-Bizet:2019eaf}, we have chosen to include a $2\pi i n_0$ shift in the definition of the chemical potential for $J_1+\tfrac{1}{2}R$,
where $n_0$ is an integer. This is because the index is a branched function of the fugacities $p=\rme^{-\omega_1}$, $q=\rme^{-\omega_2}$ (since the R-charges are generically non-integer), and $n_0$ can be used to label the different sheets. Also notice that the coefficient in front of $\{\mathcal{Q},\mathcal{Q}^\dagger\}$ in \eqref{eq:indexdef} is immaterial, as only states annihilated by $\mathcal{Q}$ and $\mathcal{Q}^\dagger$ contribute to the index.

The index \eqref{eq:indexdef} can be represented as a path integral for the SCFT on an $S^3\times S^1$ topology \cite{Closset:2013vra,Assel:2014paa,Closset:2014uda,Cabo-Bizet:2018ehj}, where $S^1$ is the Euclidean time circle of circumference $\beta$.
On general grounds, chemical potentials can be introduced either by twisting the identifications of the dynamical fields when going around $S^1$ via the action of the corresponding global symmetries, or by turning on suitable components for the background metric and the background gauge field that couples to the R-current. The two pictures are related by a large diffeomorphism together with a large ${\rm U}(1)_R$ gauge transformation. 
We choose to encode the variables $\omega_1,\omega_2$ in background fields;  on the other hand, we find it convenient to implement the insertion of 
\begin{equation}
(-1)^F \rme^{-2\pi i n_0\left(J_1+\frac{1}{2}R\right)} \,=\,
\rme^{-2\pi i (n_0-1)J_1-\pi i n_0 R}
\label{eq:insertion}
\end{equation}
in \eqref{eq:indexdef} as a boundary condition. When $n_0=0$, this yields the usual supersymmetric boundary conditions where both bosonic and fermionic fields are periodic around $S^1$.
When $n_0=\pm1$, the insertion \eqref{eq:insertion} just reads $\rme^{\mp \pi i R}$, meaning that when going once around $S^1$, every field satisfies thermal boundary conditions (periodic for bosons, antiperiodic for fermions), twisted by a phase depending on its R-charge. Since $(J_1+\tfrac{1}{2}R)$ commutes with the supercharge, this is also a
supersymmetric boundary condition:  fermion-boson cancellation is ensured by the fact that the R-charges of bosonic and fermionic fields in the same supermultiplet differ by one.

\subsection{One-form symmetry and 3d EFT}\label{sec:1form_sym_breaking}

We consider the Cardy-like regime of small chemical potentials,$
(\omega_1 + 2\pi i n_0)\to \pm 2\pi i,$ $ \omega_2\to0.$
After fixing $n_0=\pm1$, namely after moving to the “second sheet” of the branched function that gives the index, this is the same as taking 
\be\label{Cardysecondsheet}
\omega_1\to0\,,\qquad \omega_2\to 0\,,\qquad\qquad n_0=\pm 1\,.
\ee 
For definiteness, below we will refer to the choice $n_0=+1$.
The asymptotic expansion of the index in this regime has been extensively studied in the literature, see~\cite{Choi:2018hmj,Honda:2019cio,ArabiArdehali:2019tdm,Kim:2019yrz,Cabo-Bizet:2019osg,Amariti:2019mgp,ArabiArdehali:2019orz,GonzalezLezcano:2020yeb,Amariti:2020jyx,Cassani:2021fyv,ArabiArdehali:2021nsx} for a partial list of references, and is given by \cite{Cassani:2021fyv}
\be\label{eq:index}
\log \mathcal{I} \,\simeq\, \frac{(\omega_1+\omega_2+2 \pi i)^3}{48 \omega_1 \omega_2} \text{Tr} R^3-\frac{(\omega_1+\omega_2+2 \pi i)(\omega_1^2+\omega_2^2-4 \pi^2)}{48 \omega_1 \omega_2}\text{Tr} R+ \log |{G}|\,, 
\ee
up to exponentially suppressed terms that we will not consider. The expression depends on the SCFT data through the R-symmetry anomaly coefficients ${\rm Tr} R^3$ and ${\rm Tr} R$, as well as through the order $|G|$ of the discrete electric-type one-form symmetry group $G$ that the theory may have. The anomaly coefficients are finite in this analysis, namely the expression above holds at finite $N$.
When also taking a large-$N$ limit in a holographic SCFT, expression \eqref{eq:index} displays a large-$N$ growth and matches the action of the supersymmetric AdS$_5$ black hole both at the two derivative level \cite{Cabo-Bizet:2018ehj} and including higher-derivative corrections \cite{Bobev:2022bjm, Cassani:2022lrk}. The action \eqref{eq:index} then can be seen as a saddle-point contribution to the gravitational index.

For the $\mathcal{N}=4$ SU$(N)$ SYM theory, the logarithmic term $\log |{G}|=\log N$  in \eqref{eq:index} was first identified in \cite{GonzalezLezcano:2020yeb}
 by studying the explicit representation of the superconformal
index in terms of an integral of elliptic gamma functions.
This result has been extended to different
theories in \cite{Amariti:2020jyx,Amariti:2021ubd,ArabiArdehali:2021iwe}.
The general identification of the argument of the
logarithmic term with the order of a 
discrete electric one-form symmetry was proposed in \cite{Cassani:2021fyv}.
In the following we illustrate the latter statement in some detail.

For concreteness, consider a Lagrangian 4d $\mathcal{N}=1$ gauge theory with
gauge potential $\mathcal{A}$ and assume it has a discrete electric-type one-form symmetry group
$G$ of order $|G|$. The physical objects charged under this one-form symmetry are Wilson
loops
\begin{equation}
W_r[\gamma] \,=\, \mathrm{Tr}_r\, \mathcal{P}\exp\left(
i\oint_\gamma \mathcal{A}
\right)\,,
\label{eq:Wilson}
\end{equation}
where $r$ denotes the representation of the gauge group and $\gamma$ is a closed path in spacetime.
The one-form symmetry acts on Wilson loops by a phase multiplication:
\begin{equation}
W_r[\gamma]\, \rightarrow\, \exp\left(\frac{2\pi i\,k\,q_r}{|G|}\right) W_r[\gamma]\,,\qquad k = 1,\ldots,|G|\,,
\label{eq:Wphase}
\end{equation}
where $k$ parametrizes the elements of $G$ and $q_r$ is the charge of $W_r$ under the one-form
symmetry, see e.g.~\cite{Gomes:2023ahz}.
In particular, when the theory is $\mathcal{N}=4$ SYM with gauge group ${\rm SU}(N)$,
 Wilson lines in any representation are allowed, hence there is an electric one-form
symmetry given by the center of ${\rm SU}(N)$, i.e.\ $G=\mathbb{Z}_N$ so that $|G|=N$.
The parameter $k$ is an integer identified mod $N$ and the charge $q_r$ is the $N$-ality of the representation
$r$.

Representing the supersymmetric index as a path integral over $S^3\times S^1$, the Cardy-like regime
above can be implemented by taking the circumference $\beta$ of $S^1$ to be small.
It is then natural to reduce the 4d theory on $S^1$ and describe it via a 3d supersymmetric
effective field theory (EFT)~\cite{Cassani:2021fyv}. When doing so, the 4d one-form symmetry results in a one-form symmetry plus a zero-form symmetry in the 3d EFT,
both with the same group $G$ as the original symmetry \cite{Gaiotto:2014kfa}.
The zero-form symmetry acts like a gauge transformation $U$ twisting around $S^1$:
\begin{equation}
U( t_E+\beta,x) \,=\, \exp\left(\frac{2\pi i\,k\,q_r}{|G|}\right)U( t_E,x)\,,
\label{eq:twistedU}
\end{equation}
where $ t_E$ is the coordinate on $S^1$. This
 leaves the action invariant but is not really a gauge transformation since it is not
periodic in Euclidean time. 

The charged objects under this discrete zero-form transformation are Polyakov loops $P(x)$,
that is, Wilson loops winding around $S^1$:
\begin{equation}
P(x) \,=\, \mathrm{Tr}\, \mathcal{P}\exp\left(\int_0^\beta \diff t_E\,\mathcal{A}_{t_E}( t_E,x)\right)\,.
\label{eq:Polyakov}
\end{equation}
These are local operators in the 3d EFT. For simplicity we will assume the Polyakov loop is in the fundamental representation. 

Since the four-dimensional theory is conformal,\footnote{Moreover, the integrated Weyl anomaly vanishes in the background of interest~\cite{Cassani:2013dba}, so performing Weyl transformations does not modify the partition function.
} 
we can also regard the Cardy-like regime as a regime where $S^3$ is very large, with the strict limit corresponding to an infinite-volume limit. 
In the infinite-volume limit, $P(x)$ serves as order parameter for the spontaneous breaking of the 4d
one-form symmetry. The zero-form symmetry transformation \eqref{eq:twistedU} acts on $P(x)$ as a multiplication by the phase $\rme^{\frac{2\pi i\,k}{|G|}}$. When the one-form symmetry is spontaneously broken at infinite volume, the 3d EFT has then $|G|$ degenerate vacua. Correspondingly, $P(x)$ acquires a non-vanishing vev which is one among $|G|$ possibilities.

If we now consider a large but finite $S^3$, the $|G|$ configurations are still there, but we should sum over them. Although $P(x)$ does not vanish when evaluated at each individual saddle, its expectation value does vanish since it is proportional to the sum of all $|G|$-th roots of unity, $\sum_{k=1}^{|G|}\rme^{\frac{2\pi i\,k}{|G|}}=0$ \cite{Sundborg:1999ue,Aharony:2003sx}. This is consistent with the fact that there is no spontaneous breaking at finite volume. Moreover, the 3d EFT vacuum field configuration has a large but finite Euclidean action, $I_{\rm vac}$. This takes the same value in all $|G|$ configurations. Hence the corresponding contribution to the partition function reads $|G|\, \rme^{-I_{\rm vac}}+\ldots$, where $|G|$ appears as a multiplicative factor.

It just remains to confirm that the vacuum of the 3d EFT we are considering spontaneously breaks $G$.
 Central ingredients of the 3d EFT are the holonomies of the 4d gauge field around $S^1$, since they give rise to light scalar fields in 3d.
These are constructed by considering the restriction of  $\mathcal{A}_{ t_E} $ to the Cartan subalgebra of the gauge group, decomposed as $\sum_j\mathcal{A}_{t_E}^{(j)} H_j$, where
$\{ H_j\}$ is a basis for the Cartan subalgebra and $j =1,\ldots,$  rank of the gauge group. Then the gauge holonomies are the maximal torus elements $ \rme^{2\pi i u_j}$ parameterized by the periodic 3d scalars
\be
u_j(x) \,=\, \frac{1}{2\pi i} \int_0^\beta \diff t_E\,\mathcal{A}_{t_E}^{(j)}( t_E,x)\,.
\ee
While the scalars $u_j$ are flat at tree level, a non-trivial effective potential $V_{\rm eff}(u)$ is generated quantum mechanically in the EFT. It was found in \cite{Choi:2018hmj,Honda:2019cio,ArabiArdehali:2019tdm,Kim:2019yrz,Cabo-Bizet:2019osg,Amariti:2019mgp}
that in the limit \eqref{Cardysecondsheet}, assuming ${\rm Re}\,\frac{i}{\omega_1\omega_2} < 0$, the potential $V_{\rm eff}(u)$
 has a universal minimum at $u_j=0+O(|\omega|)$. For instance, for a 4d $\mathcal{N}=1$ theory based on the gauge group ${\rm SU}(N)$, the 3d EFT around this minimum is $\mathcal{N}=2$  ${\rm SU}(N)_N$ Chern-Simons gauge theory, with the $u_j$ belonging to 3d vector multiplets. This theory has a gapped supersymmetric vacuum at $u_j=0$. 

Evaluating the saddle-point action including the higher-order terms
 determined in \cite{Cassani:2021fyv} leads to the asymptotic expression for the index given by the first and second terms in \eqref{eq:index}. However, this is not the full story: the zero-form symmetry \eqref{eq:twistedU} acts by shifting all the holonomies as
 \be
u_j \,\to \, u_j + \frac{k}{|G|}\,,
\ee
 hence $V_{\rm eff}$ must be periodic and the theory has $|G|$ degenerate vacua at infinite volume. Summing over the saddles at finite volume gives $|G|$ times the same saddle-point contribution to the index, thus accounting for the third term in~\eqref{eq:index}.

Besides the $|G|$ minima of the effective potential with all the holonomies clumped together at the same value, one may consider configurations where smaller packages of the holonomies are distributed between the different minima, so that a subgroup of the one-form symmetry is preserved. In particular, given a divisor $C>1$ of $|G|$, it may be possible to assign equal packages of holonomies to the $C$ vacua at $u= 0,\, \frac{1}{C},\,\frac{2}{C},\,\ldots,\,\frac{C-1}{C}$. This configuration preserves a one-form symmetry of order $C$ and contributes with a $|G|/C$  multiplicity factor to the partition function (equivalently, the saddle-point action has a $\log \frac{|G|}{C}$ correction). Saddles of this form are indeed known~\cite{Cabo-Bizet:2019osg,ArabiArdehali:2019orz,GonzalezLezcano:2020yeb,Amariti:2020jyx}. 
 It would be interesting to study tunnelling effects involving these configurations, in relation with exponentially suppressed corrections to the Cardy-like formula~\eqref{eq:index}.\footnote{We thank Ohad Mamroud for a discussion on this point.}

\paragraph{The EFT of~\cite{DiPietro:2014bca}.} We can also comment on the different Cardy-like limit that is obtained by first setting $n_0=0$ in \eqref{eq:indexdef}, and then taking $\omega_1,\omega_2\to0$. This was first studied in \cite{DiPietro:2014bca}. Although the effective potential $V_{\rm eff}(u)$ takes a different form in this case, for large classes of SCFT's (such as many charge-conjugation invariant theories with ${\rm Tr}R < 0$) 
 it is still minimized in $u=0$ \cite{DiPietro:2014bca,Ardehali:2015bla,DiPietro:2016ond}. The saddle-point contribution to the index now reads $\log \mathcal{I}\sim -\frac{\pi^2}{6}\frac{\omega_1+\omega_2}{\omega_1\omega_2}{\rm Tr}R$. When applied to holographic SCFT's, ${\rm Tr}R$ does not grow with $N$ and therefore this expression cannot correspond to the action of a large black hole.
  However, we observe that since the gauge holonomies are clumped in $u=0$, the one-form symmetry is spontaneously broken, indicating that this regime should still describe a deconfined phase. It would be interesting to clarify the significance of this phase.
  
\subsection{Large-$N$ matrix model for ${\rm SU}(N)$ $\mathcal{N}=4$ SYM}\label{sec:largeNforSYM}

In view of the comparison with the gravity dual, we also discuss how the logarithmic term appears in the large-$N$
limit of the superconformal index, for finite values of the chemical potentials $\omega_1,\omega_2$. 
In the last decade, new methods have been developed to study the superconformal index at large $N$. These start from the fact that, for a Lagrangian theory, the superconformal index can be written as a matrix model whose integration variables are the (constant) gauge holonomies. Schematically, it reads
\be
\mathcal I(\omega_1,\omega_2;n_0) 
\,=\, \frac{1}{|\mathcal{W}|}\int \prod_{i} \diff u_i   \, \mathcal{Z}_{\rm vec} \mathcal{Z}_{\rm chi}\, ,
\ee
where  $|\mathcal{W}|$ is the order  of the Weyl group, while $\mathcal{Z}_{\rm vec}$ and $\mathcal{Z}_{\rm chi}$ are vector and chiral multiplet contributions, which can be written in terms of elliptic gamma functions \cite{Dolan:2008qi,Rastelli:2016tbz}. This is the expression that is obtained by supersymmetric localization of the SCFT path integral, see e.g.~\cite{Assel:2014paa}.
 The different methods used to study the index 
 include the Bethe-ansatz method \cite{Benini:2018ywd,Lanir:2019abx,Benini:2020gjh,Aharony:2021zkr,Mamroud:2022msu},
the elliptic extension of \cite{Cabo-Bizet:2019eaf,Cabo-Bizet:2020nkr},
the truncated matrix model of \cite{Copetti:2020dil,Choi:2021lbk},
as well as the saddle-point analysis of \cite{Choi:2021rxi,Choi:2023tiq}.
In the following we will refer to the elliptic extension of the matrix model, where the  real integration
variables $u$ of the original model are promoted to live on a torus, 
which considerably simplifies the study of saddle points~\cite{Cabo-Bizet:2019eaf}.\footnote{The relation between the elliptic extension
and the Bethe-ansatz method was discussed in \cite{Cabo-Bizet:2020ewf}.}

Let us focus on $\mathcal{N}=4$ SYM with gauge group ${\rm SU}(N)$.
For simplicity we also restrict to the case of equal chemical potentials, 
$\omega_1=\omega_2\equiv2\pi i\, \tau$, where $ \tau$ is complex.\footnote{The analysis can be extended to the case of unequal chemical potentials \cite{Benini:2020gjh,Choi:2021rxi}. This does not affect the degeneracy of saddles of interest here.} 
 For the ${\rm SU}(N)$ gauge group, the gauge holonomies can be parameterized as $\rme^{2\pi i u_j}$, $j=1,\ldots,N$, with the constraint $\sum_{j=1}^N u_j \in \mathbb{Z}+ \tau \mathbb{Z}$. The holonomies are assumed to take complex values on the torus with periods 1 and $ \tau$, so that we have the identifications $u_j \sim u_j+1\sim u_j +  \tau$.

A family of complex solutions to the
saddle-point equations of the elliptically-extended matrix model is given by the following uniform distribution of gauge holonomies:
\begin{equation}
u_j = \frac{j}{N}(m  \tau +n) + u_0 \,,\qquad j=1,\ldots,N \,,
\label{eq:uniform_u}
\end{equation}
where the integers $(m,n)$ label the different solutions, and $u_0$ is given by
\begin{equation}
u_0 = -\frac{N+1}{2N}(m \tau+n) + \frac{k}{N}\,,\qquad k\in\mathbb{Z}_N\,.
\label{eq:u0}
\end{equation}
The choice of the integer $k$ corresponds to acting with the $k$-th element of the $\mathbb{Z}^{(1)}_N$ transformations introduced in \eqref{eq:twistedU}. In the following, we will always assume that $m$ and $n$ are coprime.

For $(m,n)=(0,1)$, the choices of $k$
do not give different configurations as they correspond merely to a permutation of the
$u_j$, so the saddle is non-degenerate.
The large-$N$ action for this $(0,1)$ saddle is independent of $N$;  then, this saddle should correspond to pure AdS$_5$ on the gravity side.

For $m\neq0$ we encounter the black hole saddles.
The large-$N$ saddle-point action depends on $(m,n)$ as well as on the choice of $n_0$;
again we fix $n_0=1$ for definiteness.

The $(1,0)$ configuration has an action scaling like $N^2$, precisely corresponding
to the action of the supersymmetric black hole $I_{\rm BH}$ (with equal chemical potentials $\omega_1=\omega_2$).
This dominates over the other $(m,n)$ saddles in a suitable regime of chemical potentials \cite{Cabo-Bizet:2019eaf}. 
In particular, it is the dominating configuration in the Cardy-like regime 
discussed above.
In this case, the choices $k=0,\ldots,N-1$ give $N$ distinct configurations with the same action \cite{GonzalezLezcano:2020yeb,ArabiArdehali:2019tdm}, namely we have $N$ degenerate saddles. We conclude that the black hole saddle contribution to the index $\mathcal{I}$ is given by $N\,\rme^{-I_{\rm BH}}$.

What about the other saddles with $m\neq0$? 
Saddles with $m\neq0$ and $mn_0+2n=1$ (mod 3) or $mn_0+2n=2$ (mod 3)
have been identified on the gravity side either
with the same supersymmetric black-hole solution with shifted chemical potentials,
or (for $m>1$) with orbifolds of the original supersymmetric black hole \cite{Aharony:2021zkr}.
The saddles with $m\neq0$ and $mn_0+2n=0$ (mod 3) are an exception, 
as they do not carry an $\mathcal{O}(N^2)$ entropy; 
their gravitational interpretation is not clear.\footnote{Note that this includes the saddle $(m,n)=(1,0)$ with $n_0=0$, 
that is, the one picked by the Cardy-like limit for the $n_0=0$ index.}  

These $m \neq 0$ configurations generically completely break the one-form symmetry that acts as in \eqref{eq:u0}. However, the case where $\mathrm{gcd}(m,N)\equiv p\neq 1$ is special, as some of the configurations generated by the different choices of $k$ are actually equivalent up to a permutation of the holonomies. 
Indeed, performing the transformation \eqref{eq:u0} with $k=N/p$ shifts all holonomies by $1/p$. For $p=\gcd(m,N)$, this shift maps the set of holonomies in \eqref{eq:uniform_u} into itself, so that it does not generate a new configuration. Equivalently, the $N$ possible values of $k$ are identified in groups of $p$.
This is particularly clear when $m$ divides $N$, namely when $p=m$. In this case, the $N$ holonomies distributed along the line $m \tau+n$ organize themselves into $m$ identical groups of $N/m$ holonomies. 
A shift with $k=N/m$ maps each group into another one, leaving the unordered set of holonomies unchanged. 

\begin{figure}[h]
\centering
\includegraphics[width=0.6\textwidth]{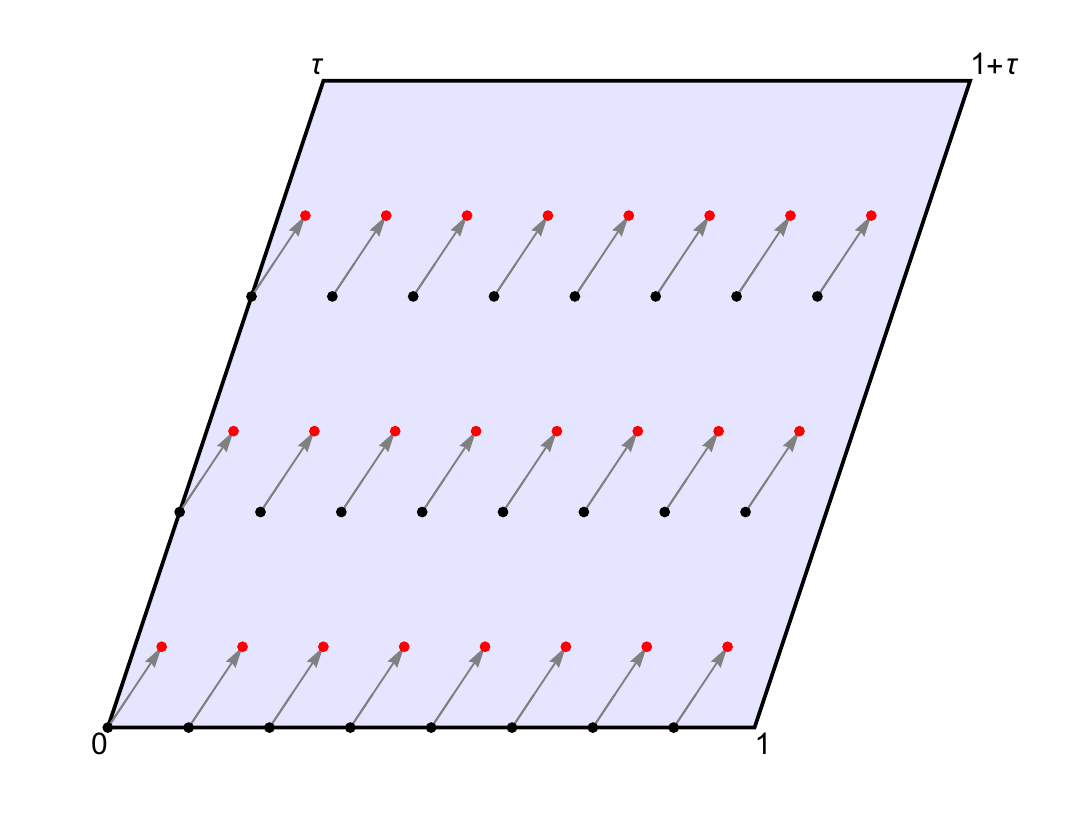}
\caption{Action of the shift \eqref{eq:shift_dual_cycle} along the cycle $\check m \tau+\check n$, dual to $m \tau+n$, on the holonomy configuration for $(m,n,\check m,\check n,N,k)=(8,3,3,1,24,1)$. The transformation maps the original distribution to a distinct but degenerate configuration, while preserving the relative arrangement of the eigenvalues on the torus.}
\label{fig:figure1}
\end{figure}

However, the action of the one-form symmetry specified in \eqref{eq:u0} does not exhaust the possible translations of the holonomies on the torus. Namely, it does not capture the full degeneracy of $(m,n)$ saddles when $m$ and $N$ have common factors. One may also consider transformations along the other cycle of the torus,
\begin{align}
    u_j \to u_j+\frac{k'}{N}\, \tau\,, \qquad k'\in\mathbb{Z}_N\,.
\end{align}
For the $(1,0)$ saddle these transformations do not generate new configurations, since a shift by $\frac{ \tau}{N}$ simply permutes the eigenvalues distributed along the $ \tau$-cycle. More generally, for the $(m,n)$ saddle, the holonomies are invariant, as an unordered set, under
\begin{align}
    u_j\to u_j +\frac{j_0}{N}\,(m  \tau+n)\,, \qquad j_0 \in \mathbb{Z}_N\,,
\end{align}
because this transformation merely shifts the eigenvalue label $u_j \to u_{j+j_0}$.
Thus, translations along the cycle $m \tau+n$ act trivially on the physical holonomy configuration.
To characterize the inequivalent configurations, we choose integers $(\check m,\check n)$ such that $n \check m-m \check  n=1$. Such integers always exist because $\mathrm{gcd}(m,n)=1$ by assumption. The cycle $\check m \tau+\check n$ is then dual to $m \tau+n$.  Different choices of $(\check m,\check n)$ differ by integer multiples of $(m,n)$ and therefore generate the same action on the space of inequivalent holonomy configurations. The inequivalent configurations can then be generated by
\begin{align}\label{eq:shift_dual_cycle}
    u_j\to u_j +\frac{k}{N}\,(\check m  \tau+\check n)\,, \qquad k \in \mathbb{Z}_N\,.
\end{align}
An explicit example of the action of this transformation on the holonomy distribution is shown in Figure~\ref{fig:figure1}.

Since the parameter $k$ takes values in $\mathbb{Z}_N$, there are $N$ inequivalent configurations of the holonomies, and the $(m,n)$ saddle contribution to the index $\mathcal{I}$ is given by $N\,\rme^{-I_{(m,n)}}$, in accordance with the results of~\cite{Aharony:2021zkr}.
 We also refer to~\cite{Cabo-Bizet:2021jar} for a discussion of the  one-form transformations acting on these $(m,n)$ saddles and their physical interpretation.

\paragraph{Polyakov loop.}
 We can also briefly comment on the order parameter
provided by the Polyakov loop \eqref{eq:Polyakov}.  
A naive calculation where this is evaluated on a chosen saddle for the gauge holonomies gives
\begin{equation}
P(m,n,k) \,=\, \sum_{j=1}^N \rme^{2\pi i u_j}
\,=\, \rme^{2\pi i k/N}\,\frac{z^N-z^{-N}}{z-z^{-1}}\,,\qquad\quad z \,=\, \rme^{i \pi\frac{m  \tau+n}{N}}\,.
\label{eq:Polyakov2}
\end{equation}
For simplicity, we have taken $m$ and $N$ coprime, so that it is enough to consider the one-form symmetry transformations in \eqref{eq:u0}.
 For $m=0$, we obtain $P(0,n,k)=0$ for $0<n<N$, and $P(0,N,k)=\rme^{2\pi i k/N}( -1)^{N+1}N$.
These values are the coefficients of the Haar-random distribution
characterizing complete confinement, see e.g.\ \cite{Hanada:2023rlk}. 
For $m\neq0$, the Polyakov loop takes a non-vanishing
value at each individual saddle labelled by $k$; the different choices of $k$ correspond to choosing one of the $N$-th roots of unity as multiplicative factor. However, the expectation value of the Polyakov loop is given by the sum over all saddles. 
 Even if the dominating contribution is given by a specific choice of $m,n$, we should still sum over $k$; this gives the sum over all roots of unity, which vanishes. Hence we conclude that $\langle P\rangle=0$ also in this regime, in agreement with the fact that there is no spontaneous breaking in finite volume.

Therefore, even if the expectation value of the Polyakov
loop vanishes  both in the unique $(0,1)$ saddle and in the degenerate family of $(1,0)$  saddles, 
the reason why it is so is quite different.

\subsection{More general SCFT's}

We next discuss how the arguments above extend to more general Lagrangian SCFT's, as well as to non-Lagrangian theories. In these cases, the one-form symmetry is not the same as the center of the gauge group.

\paragraph{More D3-brane theories.} Besides the ${\rm SU}(N)$ $\mathcal{N}=4$ SYM theory, there are infinitely many other theories
with an analogous electric one-form symmetry $G = \mathbb{Z}_N$, that gives 
black-hole saddles with degeneracy $|G|=N$.
 These theories are $\mathcal{N}=1$ quiver gauge theories flowing to a superconformal fixed point
describing the low-energy dynamics of a stack of $N$ D3-branes probing the tip of a Calabi–Yau
cone, and having a holographic dual description in terms of type IIB string theory on
$\mathrm{AdS}_5\times {\rm SE}_5$, where ${\rm SE}_5$ is the Sasaki–Einstein base of the cone.
For instance, one may consider the infinite $Y^{p,q}$ toric quivers.
 These include $\mathbb{Z}_{2p}$ orbifolds of $\mathcal{N}=4$ ${\rm SU}(N)$
SYM (given by $Y^{p,p}$) as well as the Klebanov–Witten model,
dual to type IIB string theory on $\mathrm{AdS}_5\times T^{1,1}$,
and its orbifolds (given by $Y^{p,0}$).

The structure of the quiver in the toric class is represented by a graph made of
nodes connected by arrows. The nodes represent ${\rm SU}(N)$ gauge factors,
while an arrow stemming from node $a$ and reaching node $b$ represents a bifundamental chiral
superfield transforming in the fundamental representation of $a$ and the antifundamental of $b$.
The theory may also include adjoint chiral superfields, represented by arrows with
both ends attached to the same node. In order to see that the electric one-form symmetry
group of such theories is $G = \mathbb{Z}_N$, we should recall that in the presence of matter
fields the electric one-form symmetry is the subgroup of the center of the gauge group
that leaves the matter fields invariant \cite{Gaiotto:2014kfa}.
Here the center is $(\mathbb{Z}_N)^{\nu}$ and its action
on a matter field connecting two nodes $a$ and $b$ (with $b=a$ for fields in the adjoint)
yields a phase $\rme^{2\pi i \frac{n^a - n^b}{N}}$,
where $n^a,n^b$ are integers going from 1 to $N$ parameterizing the element of the corresponding copy of $\mathbb{Z}_N$.\footnote{
Fields in generic representations of two connected nodes transform under the center, 
picking up a phase $
\exp\!\left( \frac{2\pi i}{N}\,( q^a n^a + q^b n^b ) \right)$ 
where \(q^a\) and \(q^b\) are the charges of their representations under the \(a\)-th and 
\(b\)-th copies of \(\mathbb{Z}_N\) in the center. The charges correspond to the 
\(N\)-ality of the representations; in particular,
$
q = 1$ for the fundamental, $q = N-1$ for the anti\text{-}fundamental, $q = 0$ for the adjoint.
}
In order to have invariance we need
$n^a=n^b$ for all choices of $(a,b)$ such that there is a bifundamental field connecting the
corresponding nodes in the quiver.
Given that every pair of nodes can be connected through a path, not necessarily direct,
of bifundamental fields, this ultimately means
that there is only one free $n_a$,
hence only the diagonal $\mathbb{Z}_N$ in the $(\mathbb{Z}_N)^{\nu}$ center survives
as one-form symmetry. This agrees with~\cite{Amariti:2021ubd}.

The large-$N$ limit of the superconformal index for these quiver theories has been
studied in \cite{Lanir:2019abx,Cabo-Bizet:2020nkr,Benini:2020gjh,Colombo:2021kbb,Choi:2023tiq}.
It has been found that there exist $(m,n)$ saddles
analogous to the ones discussed in section~\ref{sec:1form_sym_breaking} for $\mathcal{N}=4$ SYM:
the gauge holonomies associated with different nodes take the same values,
$u^{(a)}_j = u_j$ for all nodes $a=1,\ldots,\nu$,
with $u_j$ being given by \eqref{eq:uniform_u}.
The $\mathbb{Z}_N$ one-form symmetry acts on the gauge holonomies in
the same way as discussed in section~\ref{sec:1form_sym_breaking}, and Polyakov loops associated with different
gauge nodes all take the same values.
Again, the $(0,1)$ saddle carries no $\mathcal{O}(N^2)$ entropy and
is non-degenerate, while the $(1,0)$ saddle carries the black-hole entropy and has $N$-fold degeneracy.

Further theories having a $\mathbb{Z}_N$ electric one-form symmetry that also leads to $N$-degenerate
black-hole saddles are the non-toric quivers whose index was studied in \cite{Amariti:2021ubd},
as well as the $\mathcal{N}=1$ Leigh-Strassler deformation of ${\rm SU}(N)$ $\mathcal{N}=4$ SYM theory,
dual to the Pilch–Warner $\mathrm{AdS}_5$ solution of type IIB supergravity on a deformed $S^5$,
studied in \cite{Amariti:2020jyx}. 

For ${\rm SO}$ and ${\rm USp}$ groups the logarithmic terms in the respective index saddles
have been given in \cite{Amariti:2020jyx,Amariti:2021ubd},
and again they agree with the order of the one-form symmetry.

\paragraph{Non-Lagrangian theories from M5 branes.} Although our discussion so far has been based on theories with a Lagrangian,
 the result that existence of an electric one-form symmetry $G$ yields a   $|G|$-degeneracy of black hole saddles in the superconformal index is expected to hold for non-Lagrangian theories as well. This would be consistent with the observation that the Cardy-like expansion~\eqref{eq:index} is entirely fixed by symmetries and anomalies (in fact the first line of~\eqref{eq:index} can be derived without any reference to a Lagrangian by integrating the relevant
anomaly polynomial \cite{Ohmori:2021dzb,Cassani:2024tvk}).

For non-Lagrangian theories whose circle reduction admits a Lagrangian description, one should be able to identify the local operators in the 3d EFT which are charged under the discrete zero-form symmetry $G$ originating from the reduction of the 4d one-form symmetry, and then apply the same arguments presented above.
Among these theories one finds those describing the low-energy dynamics of $N$ M5-branes wrapping a Riemann surface of genus $g$. The analysis of~\cite{Tachikawa:2013hya} shows that (at least in the case with no punctures) the electric one-form symmetry is $\mathbb{Z}_{N^g}$, hence we expect $N^g$ degenerate black hole saddles in the associated superconformal index. 
In particular, this applies to the $\mathcal{N}=2$ \cite{Gaiotto:2009gz} and $\mathcal{N}=1$ \cite{Benini:2009mz,Bah:2012dg} non-Lagrangian SCFT's dual to the ${\rm AdS}_5$ solutions of eleven-dimensional supergravity found in \cite{Maldacena:2000mw,Bah:2012dg}.
Below we will see how holography supports this conclusion.

\section{Black hole saddle in the Cardy-like limit}\label{sec3}

The asymptotic expression \eqref{eq:index} for the SCFT index 
is derived in a Cardy-like regime of small chemical potentials, $\omega_{1,2}\!\to\!0$. These variables specify how $S^3$ is twisted over $ S^1$ in the topologically $S^3\times S^1$ background, and the Cardy-like regime can be realized by taking the radius of $S^1$  much smaller than the radius of $S^3$. Since the theory is conformal, one can equivalently take the radius of $S^3$ much larger than the radius of $S^1$, so that $S^3$ decompactifies to $\mathbb{R}^3$ in the strict limit. 
Since the  $\log |G|$ term in \eqref{eq:index} does not depend on  chemical potentials,  one can study it in the limit, as discussed in the previous section.

In this section we discuss a realization of the Cardy-like regime on the gravity side.
This is obtained by taking the boundary $S^3$ of the asymptotically $\mathrm{AdS}_5$ spacetime to be very large.
When applied to the supersymmetric non-extremal AdS$_5$ black hole solution, this also corresponds to taking a large horizon radius, such that the local geometry near the outer horizon becomes approximately flat and the relevant topological structure reduces to that of $\mathbb{R}^3\times D_2$, where $D_2$ denotes the disc capping the shrinking cycle $S^1$. The limiting configuration is a black brane solution matching (via a double Wick rotation) the topological soliton with $\mathbb{R}^3\times S^1$ boundary presented in~\cite{Anabalon:2021tua}.

\subsection{General charged rotating black hole solution}

We consider the general non-extremal charged
and doubly-rotating black hole solution of minimal five-dimensional gauged supergravity found in~\cite{Chong:2005hr}.
It solves the Einstein-Maxwell equations derived from the Lagrangian: 
\begin{align}
\mathcal{L} = (R + 12g^2)\,\star1 - \tfrac{2}{3 g^2}\,F\wedge \star F
   + \tfrac{8}{27 g^3}\,F\wedge F\wedge A\,,
\end{align}
where $A$ is an Abelian gauge field (with the normalization adopted in \cite{Cabo-Bizet:2018ehj}) and $F=\diff A$.
In asymptotically static coordinates, the metric and gauge field read
\begin{align}\label{CCLP}
    \diff s^2\,=\,& -\frac{\Delta_{\theta}((1+g^2 r^2) \rho^2 \diff t +2q\nu) \diff t }{\Xi_a \Xi_b \rho^2}+\frac{2 q \nu \omega}{\rho^2}+\frac{f}{\rho^4} \left( \frac{\Delta_{\theta} \diff t}{\Xi_a \Xi_b}-\omega\right)^2+\frac{\rho^2 \diff r^2}{\Delta_r}+\frac{\rho^2 \diff \theta^2}{\Delta_{\theta}} \nn \\
    & +\frac{r^2+a^2}{\Xi_a}\sin^2\theta \diff \phi^2+\frac{r^2+b^2}{\Xi_b}\cos^2\theta \diff \psi^2\,,\nn\\[2mm]
    A\,=\,& \frac{3g q}{2\rho^2}\left(\frac{\Delta_{\theta} \diff t}{\Xi_a \Xi_b}-\omega\right)\,,
\end{align}
where
\begin{align}
\nonumber &\nu=b \, \sin^2\theta \diff \phi+a \, \cos^2\theta \diff \psi, \qquad \qquad\,\,\omega =a \, \sin^2\theta \frac{\diff \phi}{\Xi_a}+b \, \cos^2\theta \frac{\diff \psi}{\Xi_b}\,,\\[1mm]
\nonumber &\Delta_{\theta}=1-a^2 g^2 \cos^2 \theta-b^2 g^2 \sin^2 \theta, \qquad \!\Delta_r=\frac{(r^2+a^2)(r^2+b^2)(1+g^2 r^2)+q^2+2abq}{r^2}-2m \\[1mm]
\nonumber &\rho^2=r^2+a^2 \cos^2 \theta+b^2 \sin^2 \theta, \qquad\qquad\!\! \Xi_a=1-a^2 g^2\,,\qquad \Xi_b=1-b^2 g^2\,,\\[1mm]
&f=2m\rho^2-q^2+2abqg^2 \rho^2\,.
\end{align}
The coordinates $\phi,\psi$ are $2\pi$ periodic, while 
$\theta \in [0,\pi/2]$.
The solution depends on four parameters: a mass parameter $m$, an electric charge parameter $q$, and two rotational parameters $a$, $b$. The function $\Delta_r$ determines the radial structure. We denote by $r_+$  its largest positive root, providing the position of the event horizon. The spacetime is asymptotically  AdS$_5$, with AdS inverse radius equal to $g$. We will set $g=1$ from now on. The generator of the event horizon is the Killing vector $V=\frac{\partial}{\partial t}+\Omega_1 \frac{\partial}{\partial \phi}+\Omega_2 \frac{\partial}{\partial \psi}$, with:
\begin{equation}
    \Omega_1=\frac{a(r_+^2+b^2)(1+r_+^2)+bq}{(r_+^2+a^2)(r_+^2+b^2)+abq}, \qquad \Omega_2=\frac{b(r_+^2+a^2)(1+r_+^2)+aq}{(r_+^2+a^2)(r_+^2+b^2)+abq}\,.
\end{equation}
The inverse temperature is given by:
\begin{align}
    \beta\,=\,\frac{2 \pi r_+ [(r_+^2+a^2)(r_+^2+b^2)+abq]}{r_+^4[1+(2r_+^2+a^2+b^2)]-(ab+q)^2}\,,
\end{align}
while the electrostatic potential on the horizon is:
\begin{align}
    \Phi\,=\,\frac{3 q r_+^2}{2\left((r_+^2+a^2)(r_+^2+b^2)+a b q\right)}\,.
\end{align}

We continue by briefly reviewing the analysis of~\cite{Cabo-Bizet:2018ehj}.
The solution satisfies the supersymmetry condition if the parameters satisfy the relation:
\begin{equation}\label{susy}
    q\,=\,-(a-ir_+)(b-ir_+)(1-ir_+)\,.
\end{equation}
This introduces a complexification, regardless of whether one makes the Wick rotation to Euclidean time or not. The complexification can be avoided by taking the extremal limit $r_+\to r_*$, where $r_*=\sqrt{a+b+ab}$ is the extremal horizon radius, on top of the supersymmetric limit. However we will not do so, hence we will work with the complex supersymmetric non-extremal solution in the following. 
After imposing \eqref{susy}, the chemical potentials read:
\begin{align}
\nonumber \beta &=
-\,\frac{2\pi\,(a - i r_+)(b - i r_+)\,(r_*^{2} + i r_+)}
{(r_*^{2} - r_+^{2})\left[\,2(1+a+b)r_+ + i\,(r_*^{2} - 3 r_+^{2})\,\right]} \,, \\[6pt]
\Omega_1 &=
\frac{(r_*^{2} + i a r_+)(1 - i r_+)}
{(r_*^{2} + i r_+)(a - i r_+)}, \,\,\,\,\,\,\,\,\,\,\,\, \Omega_2 =
\frac{(r_*^{2} + i b r_+)(1 - i r_+)}
{(r_*^{2} + i r_+)(b - i r_+)} \,, \\[6pt]
\nonumber \Phi &=
\frac{3 r_+ (r_+ + i)}
{2 (r_*^{2} + i r_+)} \,.
\end{align} 
The redefined chemical potentials which naturally appear in a supersymmetric partition function are given by:
\begin{align}\label{chempot}
\nonumber \omega_1 \,&=\, \beta \left(\Omega_1-1\right) \,=\,
\frac{2\pi\,(a - 1)(b - i r_+)}
{2(1 + a + b) r_+ + i\,(r_*^{2} - 3 r_+^{2})} \,, \\[6pt]
\omega_2 \,&=\,\beta \left(\Omega_2-1\right) \,=\,
\frac{2\pi\,(b - 1)(a - i r_+)}
{2(1 + a + b) r_+ + i\,(r_*^{2} - 3 r_+^{2})} \,, \\[6pt]
\nonumber \varphi \,&=\, \beta \left(\Phi-\frac{3}{2}   \right) \,=\,
\frac{3\pi\,(a - i r_+)(b - i r_+)}
{2(1 + a + b) r_+ + i\,(r_*^{2} - 3 r_+^{2})} \,,
\end{align}
and satisfy the constraint
\begin{align}\label{susyconstraint}
    \omega_1+\omega_2-2\varphi=2\pi i\,.
\end{align}
An equivalent solution also exists, where all $i$ factors in \eqref{susy}--\eqref{susyconstraint} are replaced by $-i$. 

The constraint \eqref{susyconstraint} arises from imposing the correct periodicity condition on the Killing spinor in the bulk, together with regularity of the gauge field $A$ at the horizon.

The on-shell action has been matched to the field theory expression \eqref{eq:index} even  including four-derivative terms, namely both at leading and first subleading order in a large-$N$ expansion~\cite{Bobev:2022bjm,Cassani:2022lrk}. Its two-derivative expression is:
\begin{align}\label{susyaction}
    I\,=\,\frac{2 \pi}{27 G_5} \frac{\varphi^3}{\omega_1 \omega_2}\,.
\end{align}

Taking the limit $r \to \infty$ the metric in \eqref{CCLP} takes the form:
\begin{align}\label{bdrymetric}
    \diff s^2=\frac{\diff r^2}{r^2}+r^2\left( -\frac{\Delta_{\theta}}{\Xi_a \Xi_b} \diff t^2+\frac{\diff \theta^2}{\Delta_\theta}+\frac{\sin^2 \theta}{\Xi_a} \diff \phi^2 +\frac{\cos^2 \theta}{\Xi_b} \diff \psi^2\right)\,.
\end{align}
The topology of the boundary is $\mathbb{R}\times S^3$. This can be made manifest by performing, before taking $r \to \infty$,  the bulk change of coordinates $(r,\theta) \to (\hat{r},\hat{\theta})$ such that:
\begin{align}
    \Xi_a \, \hat{r}^2 \sin^2 \hat\theta=(r^2+a^2) \sin^2 \theta\,, \qquad \Xi_b \, \hat{r}^2 \cos^2 \hat\theta=(r^2+b^2) \cos^2 \theta\,,
\end{align}
which allows to rewrite the boundary metric of \eqref{bdrymetric} as:
\begin{align}
    \diff s_{\text{bdry}}^2 \,=\,\frac{\Delta_{\theta}}{\Xi_a \Xi_b} \left(-\diff t^2 +\diff \hat\theta^2 +\sin^2 \hat\theta \,\diff \phi^2 +\cos^2 \hat\theta \, \diff \psi^2\right)\,.
\end{align}

Regularity of the general solution \eqref{CCLP} solution imposes a twisted identification of the coordinates while going around the Euclidean time circle,
\begin{align}\label{eq:identifications}
    ( t_E\,,\,\phi\,,\,\psi)\, \sim \,( t_E +\beta\,,\,\phi-i \Omega_1 \beta\,,\,\psi-i\Omega_2 \beta)\,.
\end{align}
Alternatively, one can perform the change of coordinates:
\begin{align}
     t_E=\hat  t_E\,, \qquad \phi=\hat \phi-i \Omega_1 \hat t_E\,, \qquad \psi=\hat \psi-i \Omega_2 \,\hat t_E\,,
\end{align}
which untwists the identifications:
\begin{align}
    (\hat t_E\,,\,\hat\phi\,,\,\hat\psi)\, \sim \,(\hat t_E +\beta\,,\,\hat\phi\,,\,\hat\psi)\,.
\end{align}
After this transformation, the boundary metric \eqref{bdrymetric} reads, up to a conformal factor:
\begin{align}\label{eq:hatted_coords}
    \diff s_{\text{bdry}}^2=\diff \hat{ t}_E^2 +\diff \hat\theta^2 +\sin^2 \hat\theta \,(\diff \hat\phi-i\Omega_1\,\diff \hat t_E)^2 +\cos^2 \hat\theta \, (\diff \hat\psi-i\Omega_2\, \diff \hat  t_E)^2.
\end{align}
The boundary geometry thus directly encodes the chemical potentials $i \Omega_1$ and $i \Omega_2$. Similarly, the electric chemical potential $\Phi$ can be extracted from the asymptotics  of the gauge field in \eqref{CCLP}. To this end, one should add a flat contribution,
\begin{align}
    A\,&=\, \frac{3 q}{2\rho^2}\left(\frac{\Delta_{\theta} \diff t}{\Xi_a \Xi_b}-\omega\right)-\Phi\, \diff t\,,
\end{align}
chosen such that the regularity condition $\iota_V A|_{r=r_+}=0$ is satisfied. With this prescription, the boundary value of the gauge field after Wick-rotation is:
\begin{align}
    A|_{r\to\infty}\,=\, i\Phi \,\diff  t_E\,.
\end{align}

At this point we impose the supersymmetry condition~\eqref{susy}. This introduces an additional complexification if $r_+$ is assumed real and different from the extremal value $r_*$, so the metric is always complex in this case.

\subsection{Scaling limit}

We now implement our limiting procedure. We start from the supersymmetric non-extremal black hole solution reviewed above, where the parameters satisfy~\eqref{susy}.
We consider a regime where the horizon radius $r_+$ is large. Namely, we introduce a small parameter $\epsilon$,  redefine
\be\label{rplus_rescaling}
r_{+}\rightarrow \frac{\varrho}{\epsilon}\,,
\ee
and send $\epsilon\to 0$ while keeping  $\varrho,\,a,\,b$ fixed. Expanding in powers of $\epsilon$, the metric and gauge field contain divergent terms. 
The black hole thermodynamic quantities also inherit an expansion in $\epsilon$.  In particular, the inverse temperature and chemical potentials  in \eqref{chempot} read:
\begin{equation}\label{eq:betapotentials}
\begin{aligned}
    \beta &=\frac{2\pi}{3 \varrho}\,\epsilon + \mathcal{O}(\epsilon^2)\,, \qquad
    \omega_1 = \frac{2\pi}{3\varrho}(a-1)\,\epsilon + \mathcal{O}(\epsilon^2)\,, \qquad
    \omega_2 = \frac{2\pi}{3\varrho}(b-1)\,\epsilon + \mathcal{O}(\epsilon^2)\,,
\end{aligned}
\end{equation}
with the expansion of $\varphi$ following from \eqref{susyconstraint}.
The fact that in this regime the chemical potentials $\omega_1,\omega_2$ as well as $\beta$ are $\mathcal{O}(\epsilon)$ indicates that we are reproducing a Cardy-like regime in the dual field theory, realized as a configuration where the size of the boundary $S^1$ is much smaller than the one of the boundary $S^3$. In thermal field theory, this would correspond to a high-temperature limit. Note that the on-shell action \eqref{susyaction} scales as $\mathcal{O}(\epsilon^{-2})$.
Starting from the expressions given in~\cite{Chong:2005hr} (see~\cite{Cabo-Bizet:2018ehj} for their supersymmetric non-extremal values), one can also check that the Bekenstein--Hawking entropy, the angular momenta $J_1,J_2$ and the electric charge all scale as $\mathcal{O}(\epsilon^{-3})$.\footnote{A comment on the scaling of the charges may be useful. The constraint~\eqref{susyconstraint} implies that the conserved charges conjugate to the chemical potentials $\omega_1$ and $\omega_2$ are the combinations $J_1 + \frac{1}{2}Q$ and $J_2+\frac{1}{2}Q$, respectively. These combinations commute with the supercharge and appear in the definition of the supersymmetric index, see~\eqref{eq:indexdef}. Clearly, they must scale as $\mathcal{O}(\epsilon^{-3})$. Instead, the individual scaling of $J_{1,2}$ and $Q$, taken separately, is not fixed by supersymmetry and depends on how $\beta$ behaves in the limit. In the limit we are considering, where $\beta\to0$, the charges scale as $\mathcal{O}(\epsilon^{-3})$, as discussed above. A different limit which has been discussed in the literature \cite{Choi:2018hmj,David:2020ems} is the extremal one, where $\beta\to \infty$. In this case one finds that $J_{1,2}=\mathcal{O}(\epsilon^{-3})$ while $Q=\mathcal{O}(\epsilon^{-2})$.}

The divergences in $\epsilon$ appearing in the metric and gauge field can be conveniently reabsorbed by the following rescaling of the coordinates, to be made together  with~\eqref{rplus_rescaling}:
\begin{equation}\label{redef}
    r\rightarrow \frac{r'}{\epsilon}\,, \qquad
     t_E\rightarrow \epsilon\,  t_E'\,, \qquad
    \theta\rightarrow \epsilon\,\theta'\,, \qquad
    \psi\rightarrow \epsilon\,\psi' \,.
\end{equation}
The metric and gauge field now remain finite for $\epsilon\to 0$ and read:
\begin{align}\label{rescaledCCLP}
\nonumber \diff s^2 \,&=\, \frac{{r'}^2}{{(1-b^2)^2}}\left[\left(1-b^2 -\frac{ \varrho^6}{{r'}^6} \right) \diff { t_E'}^2+ 2ib\, \frac{\varrho^6}{{r'}^6} \, \diff { t_E'}\, \diff  \psi' + \left(1-b^2  +b^2\frac{\varrho^6}{{r'}^6} \right)  \diff {\psi'}^2\right] \\ 
&\quad +\frac{r'^2}{1-a^2}\left( \diff {\theta'}^2+  {\theta'}^2 \diff \phi^2\right) +\frac{{r'}^4}{{r'}^6 - \varrho^6}\, \diff {r'}^2\, , \\[2mm]
\nonumber A \,&=\, \frac{3 \varrho^3}{2(1-b^2)\, {r'}^2}
\left(- \diff  t_E' + \ii \,b\, \diff \psi' \right)+\frac{3 \varrho}{2}\, \diff  t_E' \, .
\end{align}
The identifications~\eqref{eq:identifications} become: 
\be\label{eq:identificationsBis}
    ( t_E'\,,\,\phi\,,\,\psi')\, \sim \,( t_E' +\frac{2\pi}{3 \varrho}\,,\,\phi\,,\,\psi'-\frac{2\pi i b}{3 \varrho})\,.
\ee
This metric describes an asymptotically Euclidean AdS$_5$ space with non-compact boundary $\mathbb{R}^3 \times S^1$. 
Indeed,
the rescaling of the radial coordinate induces a conformal factor $\mathcal{O}(\epsilon^{-2})$ in the boundary metric, which is compensated by the rescaling of the other coordinates in \eqref{redef}. The rescaled angular coordinates $\theta',\psi'$  become non-compact in the limit. As a result, the boundary $S^3$ is decompactified to $\mathbb{R}^3$.
Since all terms in the metric responsible for rotation vanish in the limit, we are left with a static solution.
The only non-trivial ingredient that survives is the charge. 

The boundary $S^1$ smoothly shrinks to zero size in the bulk, so the topology of the solution is $\mathbb{R}^3\times D_2$. This is best seen after  removing the twist in the identification of the angular coordinates via\footnote{The factor of $\ii$ may be removed by analytically continuing the rotational parameter as $b\to \ii\, b$.}
\be\label{untwist_coords}
     t_E'= \zeta + \ii\, b\,z\,, \qquad
    \psi' = -\ii \,b\,\zeta + z\,.
\ee
Also making the transformation
\be
 \theta'\,=\, \sqrt{(1-a^{2})(x^{2}+y^{2})}\,, \qquad
    \phi \,=\, \tan^{-1}\!\left(\frac{y}{x}\right)\,,
\ee
the solution reads
\begin{equation}
\begin{aligned}\label{ARmetric}
\diff s^2 \,&=\, g^2 r'^2\left(\diff x^2 + \diff y^2 + \diff z^2 + f(r')\, \diff \zeta^2\right)
      + \frac{\diff r'^2}{g^2 r'^2 f(r')} \,, \\
f(r') \,&=\, 1 - \frac{\varrho^6}{r'^6} \,,\\[1mm]
A\,&=\, \frac{3 \varrho g^2}{2}\left(1-\frac{\varrho^2}{r'^2}\right) \diff \zeta -\frac{3i b\varrho g^3 }{2 } \,\diff z \,,
\end{aligned}
\end{equation}
where we reinstated the AdS radius $\frac{1}{g}$.  The coordinate $\zeta$ is compact and spacelike, while
$x,y,z$ span $\mathbb{R}^3$. The solution depends just on the parameter $\varrho$ (the dependence on $b$ in the gauge field can be removed by a trivial gauge transformation).
 The function $f(r')$ vanishes at 
\begin{equation}
r'= 
\varrho \,,
\end{equation}
where the $\zeta$-circle smoothly shrinks to zero size. From~\eqref{eq:identificationsBis}, \eqref{untwist_coords}, it follows that the period of this circle is fixed to
\begin{equation}
\Delta \zeta = \frac{2\pi}{3 \varrho g^2} \,.
\end{equation}
With this identification, the spacetime is completely smooth and static, with topology $\mathbb{R}^{3} \times D_2$. 
We have thus obtained the AdS$_5$ black brane \eqref{ARmetric} as a limiting case of the supersymmetric non-extremal black hole solution. This limit is a generalization to supersymmetric solutions of a decompactification discussed long ago in
\cite{Witten:1998zw} for the Euclidean Schwarzschild-AdS black hole.

The black brane solution above is the same regular soliton solution presented in \cite{Anabalon:2021tua}, up to a trivial Wick-rotation of one of the $\mathbb{R}^3$ coordinates.

Notice that, upon imposing the supersymmetry constraint \eqref{susy}, the original black hole solution \eqref{CCLP} preserves two supercharges. On the other hand, one can see that~\eqref{ARmetric} preserves four of the eight  supersymmetries of minimal gauged supergravity~\cite{Anabalon:2021tua}. Therefore, in taking the scaling limit there is supersymmetry enhancement.  

It may also be useful to compare our procedure with different limiting procedures previously appeared in the literature. 
The limit has common features with the one discussed in~\cite{Cassani:2024kjn}: both start from a supersymmetric non-extremal complex black hole solution and yield a  topological soliton after sending $\beta\to 0$. However, it differs from that as it produces a decompactified boundary, while in~\cite{Cassani:2024kjn} the boundary $S^3$ is not decompactified.

Our limit is also different from the one studied in~\cite{David:2020ems}, which starts from the supersymmetric and extremal solution ($\beta = \infty$) and sends the rotational parameters $a$ and $b$ to 1 (in units of the AdS radius). If one tries to implement this limit starting from the supersymmetric but non-extremal solution keeping $r_+$ fixed, one finds that $\beta$ remains finite (and complex).    
Hence this limit does not realize the field theory background studied in~\cite{Cassani:2021fyv}, where $\beta\to0$. 

Finally, we note that our limiting solution differs from the one that is obtained by taking the large horizon radius limit of the supersymmetric and extremal black hole solution. This was discussed in~\cite{Gutowski:2004ez} and gives a black hole that is not asymptotically AdS.

\section{Saddle degeneracy in the gravitational index}\label{sec:GravityDegeneracy}

We would like to identify the gravity dual of the $|G|$ degenerate saddles of
the superconformal index that were discussed in section~\ref{sec:index_saddles_and_1form}. The idea is that they correspond to different supersymmetric solutions to the supergravity equations of motion sharing the same on-shell action and boundary conditions. Recalling that we can move from a degenerate field theory saddle to the next one by acting with the one-form symmetry $G$, it is natural to use the same symmetry to also identify the degenerate saddles in gravity. 

A discrete one-form symmetry $G$ at the boundary can be modeled in the bulk by a gauge theory of 
 flat two-form potentials $b_2$, $c_2$, governed by the five-dimensional topological term~\eqref{csterm}.
 The $|G|$ different
bulk solutions should be characterized by different two-form configurations, related by the one-form transformations. Since the two-forms are flat, their different values do not affect the on-shell action, explaining the saddle degeneracy.

 For type IIB on $S^5$, this mechanism has essentially been described long ago in \cite{Aharony:1998qu} by considering a non-rotating, uncharged black hole solution. 
In the following, we extend the discussion to our more involved supersymmetric setup, also allowing for more  general compactifications. We will work having in mind the supersymmetric non-extremal black hole of the previous section, which has a capped $S^3\times D_2$ topology, as opposed to the extremal limit where $D_2$ becomes an infinite throat.
When discussing the infinite-volume limit, we will use the black brane solution discussed in section~\ref{sec3}. 
 This will considerably simplify our two-derivative analysis.

\subsection{The one-form symmetry in the bulk}\label{sec:one_form_in_bulk}

For definiteness, we consider type IIB supergravity on a manifold $Y_5$ admitting a Sasaki--Einstein structure, the most familiar example being $Y_5=S^5$. We choose boundary conditions such that an electric one-form symmetry is realized in the boundary SCFT (we will elaborate on this in section~\ref{sec:bdry_cond} and in appendix~\ref{app:bdry}).  
As seen in section~\ref{sec:index_saddles_and_1form}, this one-form symmetry is $G = \mathbb{Z}_N$, with rank $|G|=N$. 
The one-form symmetry shifts the gauge holonomies $u_j$ in \eqref{eq:uniform_u} describing the black hole index saddles, while leaving the saddle-point action invariant.

In this setup, the bulk field gauging the one-form symmetry is the NSNS two-form. Since the one-form symmetry transformations are discrete, we should focus on flat two-form fields in the bulk.
 Let us consider a two-surface $\Sigma_2$ filling the boundary $S^1$ and the integral $\alpha$ of a flat NSNS two-form potential $b_2$ through it,
 \be
 \alpha \,=\, \int_{\Sigma_2} b_2\,, \qquad\qquad \diff b_2 \,=\, 0\,.
 \ee
When the circle is contractible in the bulk, as in the case of non-extremal Euclidean black holes, $\Sigma_2$ has the topology of the disc $D_2$ and $\alpha$ is finite. 
 As explained in~\cite{Aharony:1998qu}, the bulk extension of the one-form symmetry  consists of the discrete set of transformations $b_2 \to b_2 + \diff \lambda_1$ shifting $\alpha$ as 
 \be\label{eq:1form_sym_action}
 \alpha \to \alpha + \frac{2\pi k}{ N}\,, \qquad k =0,1,\ldots,N-1\,.
 \ee
 For $k=N$ we have a true gauge transformation, which does nothing, and therefore the symmetry is $\mathbb{Z}_N$\,.

In order to prove the statements above, one considers the only term in the supergravity action where a flat two-form potential plays a role, namely the Chern--Simons term. The type IIB supergravity pseudo-action contains a term\footnote{There is no proper action for type IIB supergravity, since the self-duality constraint on the RR five-form needs to be imposed by hand. However, one can consider an effective model in 5d which has an action. The consistent truncation we will use later does have this property.}
\begin{equation}
S_{\rm IIB} \;\supset\; -\int B_2\wedge \frac{\diff C_2}{2\pi}\wedge\frac{ F_5}{2 \pi}\,,
\label{eq:CSterm}
\end{equation}
where $F_5$ is the self-dual five-form field strength carrying $N$ units of flux through the internal $Y_5$,
\be
\frac{1}{2\pi}\!\int_{Y_5}\! F_{5}\,=\, N\,.
\ee
Dimensional reduction on $Y_5$ leads to an effective theory containing the five-dimensional BF-type Chern-Simons term
\be\label{eq:bdc_term}
 S_{\rm CS}\,=\,\frac{N}{2\pi} \int b_2\wedge \diff c_2\,,
\ee
where $b_2$ and $c_2$ are the restrictions of the 10d NSNS and RR two-forms to the external spacetime. This term is always present at low energies since it has only one derivative.

It is now useful to consider the holographic dual of Polyakov loops.
 These are realised in the bulk as Euclidean fundamental 
 string worldsheets along $\Sigma_{2}$. Clearly, the transformation \eqref{eq:1form_sym_action} affects the path integral calculating the Polyakov loop expectation value, since the string action contains the coupling
$
\rme^{-\ii\int_{\Sigma_2}\! B_2}\,.
$
 The expectation value of $\ell$ identical loops
$ \braket{W^\ell[S^1]}$ can be computed inserting $\ell$ fundamental strings extending along the disc $\Sigma_2=D_2$, each of them contributing a factor $\mathrm{e}^{i \alpha}$ to the path integral.
Crucially, the $D_2\times S^3$ topology of an Euclidean black hole background allows for an RR instanton characterized by the integer
\begin{align}
s\,=\,\frac{1}{2\pi}\!\int_{S^{3}}\! \diff c_2\ \in\ \mathbb{Z}\,.
\end{align}
It follows that the bulk term \eqref{eq:CSterm} yields a contribution of the form
$
\rme^{- i N s\,\alpha }\,.
$
Then the path integral
takes the schematic form:
\be \label{polyakovloopvev}
\braket{W^\ell[S^1]} \ \propto\ \int_0^{2\pi} \diff \alpha \,\,\mathrm{e} ^{i (\ell-N s) \alpha}    \int D \Phi \,  \mathrm{e}^{-S[\Phi]}\,, 
\ee
where $\Phi$ denotes all fields different from the two-form potential giving rise to $\alpha$.
One concludes that the result vanishes unless  $\ell= N s$, so only worldsheet insertions that are neutral under $\mathbb{Z}_{N}$ can give a non-vanishing expectation value~\cite{Aharony:1998qu}. 
To make the sum over different contributions more explicit, we can decompose the integration over $\alpha$ into $N$ equivalent domains related by shifts of $\frac{2\pi}{N}$:
\begin{align}
\braket{W^\ell[S^1]} \quad \propto\quad & \sum_{k=0}^{N-1} \int_{\frac{2\pi k}{N}}^{\frac{2\pi (k+1)}{N}} \diff \alpha \, \,\mathrm{e}^{ i (\ell-N s)\alpha}
\int D \Phi\,\mathrm{e}^{-S[\Phi]}\nn\\[2mm]
\quad = \quad &\sum_{k=0}^{N-1} \,\mathrm{e}^{\,2\pi i (\ell-N s)\frac{k}{N}} \int_0^{\frac{2\pi}{N}} \diff \alpha\, \,\mathrm{e}^{ i (\ell-N s)\alpha} \,
\int D \Phi \,\mathrm{e}^{-S[\Phi]}\,,
\end{align}
where in the second line we have shifted $\alpha \to \alpha +\frac{2 \pi k}{N}$ in each interval, which corresponds to the action~\eqref{eq:1form_sym_action}. 
 In this expression the dependence on $k$ factorizes completely  
and gives a sum over $N$ equivalent contributions, resulting in
\be
\sum_{k=0}^{N-1} \mathrm{e}^{\,2\pi i (\ell-N s)\frac{k}{N}}
\,=\,
\begin{cases}
0, & \text{if } \ell \neq N s\,, \\
N, & \text{if } \ell = N s \,,
\end{cases}
\ee
where the $\ell \neq N s$ expression vanishes since it is the sum over all $N$-th roots of unity.

The argument above implies an unbroken  $\mathbb{Z}_{N}$ one-form symmetry. 
Note that we have been discussing the finite volume case so far, where the symmetry is always unbroken: although each individual two-form field configuration breaks it, at finite volume we are prescribed to sum over the inequivalent  configurations in the path integral, which restores the symmetry.

In particular, if $\ell=0$, namely if we are computing the gravitational partition function with no insertions, 
then the result is $N$ times the rest. This explains the black hole saddle degeneracy in the gravitational realization of the superconformal index.

\subsection{Boundary conditions}\label{sec:bdry_cond}

Before moving to the infinite-volume case, we  discuss the boundary conditions realizing the one-form symmetry of interest.

The topological term \eqref{csterm} (and \eqref{eq:bdc_term}) has the form of a BF coupling, invariant under (small) gauge transformations $b_2 \to b_2+ \diff \lambda_1,\, c_2 \to c_2+ \diff \gamma_1$. The BF coupling implies that these symmetries are not independent, but rather organize into a discrete higher-form symmetry structure. In particular, the bulk theory exhibits a $\mathbb{Z}_N^{(2)} \times \mathbb{Z}_N^{(2)}$ symmetry, generated by 2d surface operators built from the exponentials of $b_2$ and $c_2$ \cite{Brennan:2023mmt}.
Although two gauge symmetries are present in the bulk, only one can be realized as a global symmetry at the boundary. Indeed, the  BF term forbids imposing identical boundary conditions for both two-form fields, since one acts as the conjugate momentum of the other.\footnote{Here we are referring just to the one-derivative Chern--Simons term, ignoring the two-derivative terms, since it is the only term relevant for the flat two-form potentials that gauge the one-form symmetry.} Consequently, only one linear combination of the bulk two-form gauge symmetries can be realized as a global one-form symmetry of the boundary theory, while the other is gauged.

The choice of boundary conditions for the two-form potentials plays therefore a crucial role. These determine which of the one-form symmetries associated to gauge transformations of $b_2$ and $c_2$ is realized as a global one-form symmetry of the boundary theory, and which is gauged. Imposing Neumann boundary conditions leaves the corresponding field unconstrained at the boundary, allowing it to be integrated over in the path integral, whereas Dirichlet boundary conditions fix its boundary value. Depending on which two-form is fixed, the resulting boundary one-form symmetry is of electric or magnetic type. From the ten-dimensional perspective, this distinction reflects the fact that the fields $B_2$ and $C_2$ couple to fundamental strings and D1-branes, respectively. In the dual field theory, these give rise to Wilson lines and ’t Hooft lines, which are charged, respectively, under electric and magnetic one-form symmetries \cite{Bhardwaj:2023kri}. Imposing Dirichlet boundary conditions for the five-dimensional field $b_2$ fixes its boundary value and forces the charge associated with the $c_2$ gauge transformation to vanish. As a result, magnetic one-form charges are excluded, and the boundary theory exhibits a purely electric $\mathbb{Z}_N^{(1)}$ one-form symmetry. More general mixed boundary conditions can be implemented by adding suitable boundary terms to the action, allowing for a realization of a $\mathbb{Z}_P^{(1)} \times \mathbb{Z}_Q^{(1)}$ symmetry with $PQ=N$ \cite{Apruzzi:2021phx,Bergman:2022otk}. Such choices are also reflected in the global structure of the boundary gauge group, selecting either ${\rm SU}(N)$ or a discrete subgroup $ {\rm SU}(N)/\mathbb{Z}_K$.

Since our goal is to reproduce the discrete one-form symmetry of a boundary theory with gauge group ${\rm SU}(N)$, we impose Dirichlet boundary conditions for the field $b_2$, ensuring that the one-form symmetry is of electric type.  
In appendix \ref{app:bdry}, we present a detailed discussion of the admissible boundary conditions for the two-form fields, as well as of the specific choice adopted in the model based on a Sasaki--Einstein truncation to be discussed below.

\subsection{Infinite volume}

Recall that the SCFT expression~\eqref{eq:index} is derived in a Cardy-like regime of small chemical potentials, and that this can be seen as a regime where the $S^3$ in the background geometry becomes very large and decompactifies in the strict limit. In this case the one-form symmetry can be spontaneously broken and the Polyakov loop can take a vev.
The $|G|$ degenerate saddles now correspond to $|G|$ degenerate vacua for the effective theory in $\mathbb{R}^3$.

In section~\ref{sec3}, we showed how implementing a strict Cardy-like limit in the supersymmetric non-extremal black hole solution \eqref{CCLP} gives the black brane solution~\eqref{ARmetric}, where the boundary $S^3$ is decompactified to $\mathbb{R}^3$ and the five-dimensional topology is $\mathbb{R}^{3}\times D_2$. The latter solution provides a convenient framework for studying the multiplicity of black hole saddles.
Indeed, this background captures the universal topological data responsible for the $\log |G|$ contribution to the superconformal index, while avoiding the complications associated with the finite-volume geometry. At infinite volume, the instantonic contributions that played a central role in identifying the $\mathbb{Z}_N$ symmetry in the finite volume case are absent and we should consider a slightly different mechanism, also first described in \cite{Aharony:1998qu}.
Compared to the discussion in~\cite{Aharony:1998qu}, we have the additional ingredient of the electric charge, while the rotation terms in the metric are conveniently washed away in the decompactification limit. We are going to show that the charge does not affect the degeneracy of  vacua and, in particular, it does not modify the $|G|$–fold multiplicity of saddles found in the uncharged case. Although this is expected for a topological theory such as the Chern--Simons theory, we will illustrate how this is also true when including the two-derivative kinetic terms considered in~\cite{Aharony:1998qu}.\footnote{The consequences of taking the two-derivative terms into account in the description of the one-derivative topological term have been scrutinized in \cite{Belov:2004ht}. 
}

We will formulate our analysis in a specific consistent truncation of type IIB supergravity on a Sasaki--Einstein manifold $Y_5$, which has the virtue of including both the solution of interest and the relevant two-forms $b_2$, $c_2$ within a supersymmetric five-dimensional theory with a known uplift.
Within this truncated theory, we demonstrate that the quantization of the two-form reproduces the expected $N$ distinct sectors.

\subsubsection{Two-derivative analysis}\label{sec:two-ders}
We find it useful to begin with a model including  two-derivative terms, even though the core of the argument ultimately relies on the one-derivative Chern--Simons term. Indeed, the main result of this section—namely the existence of $N$ degenerate saddles—can already be obtained by restricting attention to the one-derivative sector, where the theory is purely topological. Nevertheless, including the two-derivative terms allows us to make direct contact with the analysis of \cite{Aharony:1998qu}. Moreover, including the two-derivative terms provides additional insight into the structure of the vacuum sector. Indeed, in the purely one-derivative theory the Hamiltonian vanishes identically, and the analysis reduces to a counting of states in the topological Hilbert space. Once the two-derivative terms are retained, the Hamiltonian becomes nontrivial, allowing one to count vacua as minima of a non-trivial energy spectrum. In this framework we find that the theory exhibits $N$ distinct minima of the energy, each corresponding to one of the degenerate saddles identified in the topological analysis. The  two-derivative formulation thus makes the physical interpretation of the degeneracy in terms of distinct vacua  manifest.

The model we are going to consider arises as a subsector of the consistent truncation of type IIB supergravity on a general Sasaki--Einstein structure presented in  \cite{Cassani:2010uw,Liu:2010sa,Gauntlett:2010vu}.
We discuss the derivation of this model in appendix~\ref{app:5dmodel}. Specifically, in appendix~\ref{app:truncation} we discuss the subtruncation of the action of~\cite{Cassani:2010uw} retaining the 5d two-form potentials $b_2$ and $c_2$, while in appendix~\ref{app:bdry}  we discuss the boundary conditions for these fields.
Then, in appendix \ref{app:fluct} we study the fluctuations of $b_2$ and $c_2$ around the asymptotically AdS$_5$ background $M$ in \eqref{ARmetric}.  For any field $\Phi$ appearing in the 5d action, we denote by $\vev{\Phi}$ the value taken in the solution to the equations of motion, and by $\Phi'$ the fluctuation, so that $\Phi = \vev{\Phi} + \Phi'$. Allowing for flat background values of $b_2$ and $c_2$ and focusing on fluctuations with both indices along $\mathbb{R}^3$, the resulting quadratic action contains the standard topological coupling $\hat{b}_2 \wedge \diff c_2'$ together with additional interaction terms induced by the non-vanishing background gauge field $\vev{A}$, which are absent in the AdS-Schwarzschild case. The model we obtain, truncated at quadratic order in the fluctuations, reads: 
\begin{eqnarray}
\nonumber
   &&\!\!\!\!\!\!\!\!\!\!\!S_{\rm 5d} =\frac{(2\pi)^2}{2 \kappa_5^2} \int_{M} -\frac{1}{2} \rme^{-\hat{\phi}} \rme^{\frac{16}{3} \hat{U}+\frac{4}{3} \hat{V}}(\diff b_2'-b_1'\wedge \diff \hat{A})\wedge \star (\diff b_2'-b_1' \wedge \diff \hat{A})\\[4mm]
\nonumber &&\!\!\!\!\!\!\!\!\!\!\!\!\!\!-\frac{1}{2} \rme^{\hat{\phi}}\rme^{\frac{16}{3} \hat{U}+\frac{4}{3}\hat{V}} \left[(\diff c_2'-c_1' \wedge \diff \hat{A})-\hat{C}_0(\diff b_2'-b_1' \wedge \diff \hat{A})\right] \wedge \star \left[(\diff c_2'-c_1' \wedge \diff \hat{A})-\hat{C}_0(\diff b_2'-b_1' \wedge \diff \hat{A})\right]\\[4mm]
 &&\!\!\!\!\!\!\!\!\!\!\!\!\!\! +2 \, \diff \hat{A}\wedge \left[(\hat{b}_2+b_2') \wedge \diff c_0'+(\hat{b}_0+b_0')\, \diff c_2'\right]-2K\, \hat{b}_2 \wedge \diff c_2'\,,
\end{eqnarray}
with the notation being introduced in the appendix.
Since we are interested in the fluctuations of $b_2'$ and $c_2'$ we discard all terms that do not involve these fields. The action then reduces to:
\begin{align}\label{actsimpl}
    \nonumber  S_{\rm 5d}&=\frac{(2\pi)^2}{2 \kappa_5^2} \int_{M} \rme^{\frac{16}{3} \hat{U}+\frac{4}{3} \hat{V}}\left[\left(-\frac{1}{2} \rme^{-\hat{\phi}}-\frac{1}{2} \rme^{\hat{\phi}} \hat{C}_0^2\right) \diff b_2'\wedge \star \, \diff b_2'-\frac{1}{2} \rme^{\hat{\phi}} \diff c_2' \wedge \star \,\diff c_2'+\rme^{\hat{\phi}}\hat{C}_0 \,\diff c_2' \wedge \star \,\diff b_2'\right]\\[4mm]
&+2\,\diff \hat{A}\wedge (b_2' \wedge \diff c_0'+b_0'\,\diff c_2')-2K \,\hat{b}_2\wedge \diff c_2'\,.
\end{align}
By defining:
\begin{align}
    \int_{D_2} \hat{b}_2 =\alpha
\end{align}
and integrating over the internal $D_2$, we obtain the 3d action:
\begin{align}
    \nonumber  S_{\rm 3d}&= \int_{\mathbb{R}^3} -\frac{|\tau_s|^2}{2e^2}  \diff b_2'(\mathbf{x})\wedge \star_3\, \diff b_2'(\mathbf{x})-\frac{1}{2e^2}  \diff c_2'(\mathbf{x}) \wedge \star_3 \,\diff c_2'(\mathbf{x})+\, \frac{\text{Re} (\tau_s)}{e^2} \,\diff c_2'(\mathbf{x}) \wedge \star_3 \,\diff b_2'(\mathbf{x})\\[4mm]
&+\chi \, \left[b_0'(\mathbf{x})\,\diff c_2'(\mathbf{x})-c_0'(\mathbf{x})\,\diff b_2'(\mathbf{x})\right]- \frac{N}{2\pi}  \alpha\, \diff c_2'(\mathbf{x})\,,
\end{align}
where, after setting $g=1$:
\begin{align}\label{redefinitions}
\nonumber \frac{1}{e^2}&=\frac{(2\pi)^2}{2\,\kappa_5^2 \,\text{Im}(\tau_s)}\rme^{\frac{16}{3} \hat{U}+\frac{4}{3} \hat{V}}\int_{D_2}\frac{1}{r'^3} \diff r' \,\diff \zeta=\frac{2\pi^3 }{3\varrho^3\kappa_5^2 \,\text{Im}(\tau_s)}\rme^{\frac{16}{3} \hat{U}+\frac{4}{3} \hat{V}}, \\[4mm]
\tau_s&=\hat{C}_0+i \rme^{-\hat{\phi}}, \,\,\,\,\ N=\frac{(2\pi)^3K}{\kappa_5^2} \in \mathbb{Z},\,\,\,\,\,\,\,\, \chi =\frac{(2 \pi)^2}{\kappa_5^2}\int_{D_2} \diff \vev{A}=\frac{4\pi^3}{\kappa_5^2}\,.
\end{align}
We can choose a gauge such that the only non-vanishing components of $b_2'$ and $c_2'$ are $b_{2,12}$ and $c_{2,12}$. The mixed kinetic term can be eliminated by performing the field redefinitions:
\begin{align}
\nonumber &b_{2,12}'=\frac{1}{\text{Im} (\tau_s)}\,q_1,\\
&c_{2,12}'=q_2+ \frac{\text{Re} (\tau_s)}{\text{Im} (\tau_s)} \, q_1 \,.
\end{align}
In terms of these variables, the action takes the simplified form:
\begin{align}
    S_{\rm 3d}\,&=\, \int_{\mathbb{R}^3} \left(\frac{1}{2e^2}\dot{q}^2_1(\mathbf{x})+\frac{1}{2e^2}\dot{q}^2_2(\mathbf{x})- \phi(\mathbf{x}) \,\dot{q}_1(\mathbf{x})- \rho(\mathbf{x}) \, \dot{q}_2(\mathbf{x}) \right)\diff^3 x\,,
\end{align}
where 
\begin{align}
\nonumber    &\phi(\mathbf{x})\,=\, \frac{\chi}{\text{Im}(\tau_s)} \left(c_0'(\mathbf{x})-b_0'(\mathbf{x})\, \text{Re}(\tau_s)\right)+\frac{\text{Re}(\tau_s)}{\text{Im}(\tau_s)} \frac{N \alpha}{2\pi}\,,\\
    &\rho(\mathbf{x})\,=\,-\chi  \, b_0'(\mathbf{x})+\frac{N \alpha}{2\pi}\,.
\end{align}
The canonical momenta associated with $q_1$ and $q_2$ read:
\begin{align}
 \nonumber   &\Pi_1(\mathbf{x})=\frac{\dot{q}_1(\mathbf{x})}{e^2}-\phi(\mathbf{x})\,,\\
    &\Pi_2(\mathbf{x})= \frac{\dot{q}_2(\mathbf{x})}{e^2}-\rho(\mathbf{x})\,,
\end{align}
and the Hamiltonian reads:
\begin{align}
\nonumber      \mathcal{H}\,&=\,\frac{e^2}{2} \int_{\mathbb{R}^{2}} \left[ \left( \Pi_1(\mathbf{x})+\phi(\mathbf{x})\right)^2 +\left( \Pi_2(\mathbf{x})+\rho(\mathbf{x})\right)^2 \right]\,\diff^2 x\,.
\end{align}
We consider the following basis of gauge-invariant states:
\begin{align}\label{states}
\nonumber    \Psi\,&=\,\rme^{is \int_{\mathbb{R}^2}b'_{2,12}(\mathbf{x})+it \int_{\mathbb{R}^2}c'_{2,12}(\mathbf{x})}\\
    \,&=\,\rme^{i\frac{s+t \,\text{Re} (\tau_s)}{\text{Im} (\tau_s)}\int_{\mathbb{R}^2}q_{1}(\mathbf{x})+it \int_{\mathbb{R}^2}q_{2}(\mathbf{x})}\,,
\end{align}
with $s,t \in \mathbb{Z}$. The corresponding energy density is:
\begin{align}
    \mathcal{E}(\mathbf{x})\,&=\,\frac{e^2}{2 }\left[ \left(\frac{s+t \,\text{Re} (\tau_s)}{\text{Im}(\tau_s)}+\phi(\mathbf{x})\right)^2+\left(t+\rho(\mathbf{x})\right)^2\right]\,.
\end{align}
The vacuum configurations of the theory are obtained by minimizing $\mathcal{E}(\mathbf{x})$. For the purpose of this analysis, the explicit spatial configurations of the background fields $\phi(\mathbf{x})$ and $\rho(\mathbf{x})$ in the vacuum are not essential. Instead, we focus on characterizing the set of minima of $\mathcal{E}(\mathbf{x})$ and, in particular, their multiplicity. 
Under the simultaneous shift
\begin{align}
\alpha \rightarrow \alpha+\frac{2\pi}{N}, \qquad t \to t-1\,,
\end{align}
the fields $\phi(\mathbf{x})$ and $\rho(\mathbf{x})$ transform as:
\begin{align}
\nonumber    &\phi(\mathbf{x}) \rightarrow \phi(\mathbf{x})+\frac{\text{Re} (\tau_s)}{\text{Im}(\tau_s)}\,,\\
    &\rho(\mathbf{x}) \rightarrow \rho(\mathbf{x})+1\,.
\end{align}
while leaving $\mathcal{E}(\mathbf{x})$ invariant. After $N$ such shifts, the parameter $\alpha$ changes by $\Delta \alpha=2\pi$, which corresponds to a trivial gauge transformation of $\hat{b}_2$:
\begin{align}
    \hat{b}_2 \to \hat{b}_2 + \diff \lambda_1\,, \qquad\quad \int_{\Sigma_2} \diff \lambda_1 \in 2\pi \mathbb{Z}\,.
\end{align}
We conclude that the theory has $N$ distinct vacua due to the spontaneous breaking of the $\mathbb{Z}_N$ one-form symmetry. 

\subsubsection{One-derivative argument}
We now show how the conclusion reached above can also be derived by restricting the analysis to the one-derivative sector of the action and neglecting all kinetic terms. 

In the action \eqref{actsimpl}, the only one-derivative terms are:
\begin{align}
  S_{1\rm{-der}}\,=\,\frac{(2\pi)^2}{\kappa_5^2} \int_{M} \frac{\diff \hat{A}}{2\pi}\wedge (b_2' \wedge \diff c_0'+b_0'\,\diff c_2')-K \,\hat{b}_2\wedge \diff c_2'\,.
\end{align}
Notice that the first two terms feature two derivatives, but one of them acts on $\hat{A}$ which is a fixed background field. In the low-energy limit of the theory we can consider as frozen the background fields $b_0'$ and  $c_0'$. Since these fields are fluctuations, the only meaningful choice is to take $b_0'=c_0'=0$. In the 3d theory, after using \eqref{redefinitions}, we get the following contribution: 
\begin{align} S_{1\rm{-der}}&=-\frac{N}{ 2\pi} \int_{\mathbb{R}^3} \alpha \, \diff c_2'(\mathbf{x})\,, 
\end{align} 
whose Hamiltonian is identically zero. By retaining only the one-derivative contributions, we isolate the sector of the theory that governs the vacuum properties. While this truncation discards excited states, it preserves the information required to characterize the structure of the vacuum and its degeneracy. In this limit, the conjugate momentum takes the form:
\begin{align} 
\Pi_{c_2'}=-\frac{N \,\alpha}{2\pi}\,.
\end{align}
By continuing to consider states of the form \eqref{states}, we are led to the identification:
\begin{eqnarray} 
\alpha =-\frac{ 2\pi t }{N}\,.
\end{eqnarray}
As before, the value of $\alpha$ can be shifted by $\frac{2\pi}{N}$ via the gauge transformation $t \rightarrow t-1$. This allows us to identify the discrete values of $\alpha$ that label the $N$ ground states. As expected, by reducing the theory to the one-derivative sector we reproduced the vacuum structure identified in the two-derivative analysis. 

\subsection{Generalizations}
\subsubsection{Orbifolds and $(m,n)$ saddles}
Besides the ordinary black hole, it is possible to consider the contribution of other saddles to the gravitational index, matching the $(m,n)$ saddles on the SCFT side discussed in section \ref{sec:largeNforSYM}. These gravitational saddles are given by  $\mathbb Z_m$ orbifolds of the original supersymmetric non-extremal black hole solution, restricting for simplicity to the case $\Omega_1=\Omega_2\equiv \Omega$~\cite{Aharony:2021zkr}.

In order to describe the orbifolding procedure, we start from the 10d uplift to type IIB on $S^5$ of a black hole with chemical potentials $\tilde\beta,\tilde\Omega,\tilde\Phi$,
and identifications: 
\begin{align}\label{eq:identificationsparentBH}
( t_E,\phi,\psi,\phi_1,\phi_2,\phi_3)
\sim
\left(
 t_E+\tilde\beta,\phi-i\tilde\Omega\tilde\beta,\psi-i\tilde\Omega\tilde\beta,\phi_1-\frac{2i}{3}\tilde\Phi \tilde \beta ,\phi_2-\frac{2i}{3}\tilde\Phi \tilde \beta,\phi_3-\frac{2i}{3}\tilde\Phi \tilde \beta
\right)\, ,
\end{align}
where $\phi_1$, $\phi_2$, $\phi_3$ are the three $2\pi$-periodic angles in $S^5$.
The corresponding $\mathbb Z_m$ orbifold is obtained by quotienting by the identifications
\begin{align}\label{eq:identificationsorbifold}
\nonumber ( t_E,\phi,\psi,\phi_1,\phi_2,\phi_3)
\sim
\Biggl(& t_E+\frac{\tilde\beta}{m},\phi-i\tilde\Omega\frac{\tilde\beta}{m}-\frac{2\pi n}{m},\psi-i\tilde\Omega\frac{\tilde\beta}{m}-\frac{2\pi n}{m},\\
&\phi_1-\frac{2i}{3}\tilde\Phi\frac{\tilde\beta}{m}-\frac{2\pi s}{m},\phi_2-\frac{2i}{3}\tilde\Phi\frac{\tilde\beta}{m}-\frac{2\pi s}{m},\phi_3-\frac{2i}{3}\tilde\Phi\frac{\tilde\beta}{m}-\frac{2\pi s}{m} \Biggr) \, ,
\end{align}
with $n=0,\ldots,m-1$ and $s=0,\ldots,2m-1$. Preserved supersymmetry requires
\begin{align}
2n-3s\,=\, m\pm1 \pmod{2m} \,.
\end{align}
Therefore the orbifold saddle, which has boundary conditions specified by the quantities
\begin{align}
\left(\frac{\tilde\beta}{m},\,\tilde\Omega-\frac{2\pi i n}{\tilde\beta},\,\tilde\Phi-\frac{3\pi i s}{\tilde\beta}\right)\,\equiv\,(\beta,\,\Omega,\,\Phi)\,,
\end{align}
contributes to the same gravitational index as a black hole having chemical potentials $(\beta,\Omega,\Phi)$ \cite{Aharony:2021zkr}.
 
 It is natural to ask how the one-form symmetry acts in the orbifold geometry. 
  In this saddle, the different vacua are related by shifts of the following quantity:
\begin{align}\label{alphaorbifold}
\alpha\,=\,\int_{D_2}b_2\,=\,\frac{1}{m}\int_{\tilde D_2} b_2\,=\,\frac{\tilde\alpha}{m}\,,
\end{align}
where $\tilde D_2$ is the corresponding disc in the parent geometry, whose thermal circle has length $\tilde\beta$. As in the unorbifolded black hole case, the one-form symmetry acts as $\alpha\to\alpha+\frac{2\pi k}{N}$, with $k\in\mathbb Z_N$. However, the gauge identification $\tilde\alpha\sim\tilde\alpha+2\pi$ inherited from the parent geometry implies, using \eqref{alphaorbifold}:
\begin{align}\label{gaugeidentorbifold}
\alpha\sim\alpha+\frac{2\pi}{m}\,.
\end{align}
Thus, configurations whose values of $\alpha$ differ by $2\pi/m$ are gauge equivalent.
If $\mathrm{gcd}(m,N)=1$, the transformations $\alpha \to \alpha+\frac{2\pi k}{N}$ produce $N$ different configurations contributing equally to the gravitational index. Indeed, since $m$ and $N$ have no common factors, the additional gauge identifications introduced by the orbifold action in \eqref{gaugeidentorbifold} do not relate any configurations that were distinct in the parent black-hole geometry.

The situation changes when $\mathrm{gcd}(m,N) \neq 1$. As discussed in section \ref{sec:largeNforSYM}, in this case the orbifold gauge identifications act nontrivially on the configurations generated by the real shifts of $\alpha$. Consequently, some configurations that were distinct in the parent geometry become gauge equivalent after orbifolding. 
In the case under consideration, this reduces the number of inequivalent configurations generated by $\alpha \to \alpha+\frac{2\pi k}{N}$ from $N$ to $\frac{N}{m}$. 
This reflects the picture we have found on the SCFT side.

The real shifts of $\alpha$ therefore do not by themselves reproduce the full $N$-degeneracy expected from the field-theory analysis. To recover the missing configurations, we should be including complex shifts:
\begin{align}
    \alpha \to \alpha+\frac{2\pi k}{N}+\frac{2\pi k'}{N}  \tau\,,\qquad k,k' \in \mathbb{Z}.
\end{align}
This extension is suggested directly by the field-theory description. When $\mathrm{gcd}(m,N) \neq 1$ the $N$ equivalent holonomy configurations are obtained by performing a shift along the dual cycle $\check{m}  \tau+\check n$. The additional term proportional to $ \tau$ in the transformation of $\alpha$ is the gravitational counterpart of this displacement along the dual torus cycle. The complexification of the $b_2$ field would then mirror the complexification of the holonomies in the elliptic extension of section~\ref{sec:largeNforSYM}. However, we have not been able to obtain a derivation of this complexified transformation just based on the principles of the gravitational path integral, without use of the field theory input.

\subsubsection{Multi-charge black holes}
The analysis presented above extends straightforwardly to multi-charge supersymmetric black hole solutions. 
  A natural generalization of the solution discussed in this paper is provided by solutions with three independent electric charges and two angular momenta, which are realized in $U(1)^3$ gauged $\mathcal{N}=2$ supergravity coupled to two vector multiplets \cite{Gutowski:2004yv,Cvetic:2004ny,Wu:2011gq}.  This  theory admits an embedding in type IIB supergravity on $S^5$, or an orbifold thereof.
In this setting, one can still implement the gravitational Cardy-like limit discussed in section~\ref{sec3}, now starting from a  supersymmetric non-extremal multi-charge solution such as the one discussed in~\cite{Cassani:2019mms} for equal angular momenta. The resulting configurations describe supersymmetric multi-charge black branes with $S^1\times \mathbb{R}^3$ boundary, as in \cite{Anabalon:2024che}.
 The charge parameters $Q_I$ appearing in these gravitational solutions do not modify the topological coupling relevant for the vacuum degeneracy. Consequently, the saddle degeneracy is unchanged. The two-derivative analysis presented in section~\ref{sec:two-ders} also goes through in a similar way. 
 
 This conclusion is matched in the dual SCFT, that is $\mathcal{N}=4$ ${\rm SU}(N)$ SYM or one of its orbifold theories: turning on additional chemical potentials conjugate to the electric charges in the superconformal index $\mathcal{I}$ does not affect the one-form symmetry.

\subsubsection{M-theory}
Although the results above have been presented having in mind the uplift of five-dimensional supergravity to type IIB supergravity, they also apply to different setups upon making suitable modifications. For instance, we can consider the M-theory origin of the five-dimensional solutions in section~\ref{sec3} by employing a consistent truncation of M-theory to five-dimensional $\mathcal N = 2$ minimal gauged supergravity~\cite{Gauntlett:2006ai}. In particular, this type of geometries arises from the backreaction of $N$ M5-branes wrapped on a Riemann surface $\Sigma_g$ of genus $g$, with explicit AdS$_5$ solutions having been found in~\cite{Maldacena:2000mw,Bah:2012dg}. Recall that, in this case, we have $|G|=N^g$.
This supersymmetric gravitational solution can be holographically interpreted in terms
of the large-$N$ limit of the class $\mathcal{S}$ SCFT's that arise
from the twisted compactification of the six-dimensional $(2,0)$ theory on $\Sigma_g$. Since these theories are non-Lagrangian, the global structure of their higher-form symmetries is more subtle than in the type IIB constructions previously discussed. In particular, while in the Lagrangian case one can directly read off the rank of the discrete electric one-form symmetry from the center of the gauge group, in the non-Lagrangian case the identification is less straightforward and uses the brane origin of the theory. In the present case, one has $|G|=N^g$.
We can however apply the same gravitational argument used above: by computing the degeneracy of bulk saddles we infer the discrete one-form symmetry of the dual field theory, thus bypassing the absence of a Lagrangian description.

In the earlier IIB discussion we analyzed the vacua of the effective 3d theory both at the level of one-derivative terms—which control the saddle degeneracy—and at the level of two-derivative terms—which confirm their stability. In the present M-theory case, we restrict to the one-derivative sector, since a consistent truncation preserving a topological coupling of the form \eqref{csterm} is not known. Reducing M-theory on the internal space $M_6$ yields a five-dimensional topological action of the form \cite{Bah:2020uev}:
\begin{align}\label{mthredact}
    S=\frac{1}{2\pi}\int_{M}\left( K \,c_3 \wedge \diff A_1+N\, b_{2,i} \wedge \diff \tilde{b}^i_2\right)\,,
\end{align}
where $M$ denotes the external spacetime and the fields $A_1,b_{2,i},\tilde{b}_{2,i},c_3$ arise from expanding the eleven-dimensional three-form $C_3$ on a basis of cohomology classes of $\Sigma_g$. The index $i=1,\dots,g$ labels the pairs of one-cycles of $\Sigma_g$.\footnote{In detail, $\Sigma_g$ has $b^1(\Sigma_g)=2g$ one-cycles. These naturally split into $g$ $A$-cycles and $g$ $B$-cycles, giving rise to $g$ two-forms $b_{2,i}$ and $g$ two-forms $\tilde{b}_{2}^i$, respectively.}

In the bulk, the presence of a term like $\frac{N}{2\pi}\int_{M} b_2  \wedge \diff \tilde b_2$ implies the existence of a discrete gauge symmetry $\big( \mathbb{Z}^{(1)}_N \times \mathbb{Z}^{(1)}_N\big)^g$.
The term $\frac{K}{2\pi} \int_M \, c_3 \wedge \diff A_1$ signals the presence of a discrete gauge symmetry $\mathbb{Z}^{(0)}_K \times \mathbb{Z}^{(2)}_K$. Depending on the boundary conditions chosen for $A_1$ and $c_3$, the boundary theory exhibits either a $\mathbb{Z}^{(0)}_K$ symmetry (Dirichlet b.c.\ on $A_1$) or a $\mathbb{Z}^{(2)}_K$ symmetry (Dirichlet b.c.\ on $c_3$). However, the term of interest for us is the second in \eqref{mthredact}, which guarantees a $\big(\mathbb{Z}^{(1)}_N \big)^{g}$ symmetry in the boundary theory, which is electric or magnetic depending on whether one imposes Dirichlet or Neumann boundary conditions on the two-forms $b_{2,i}$ and $\tilde b_2^i$. The physical objects charged under the one-form symmetry are M2-branes wrapping a cycle in $\Sigma_g$, forming a loop at the AdS boundary and extending in the bulk. In particular, the holographic counterpart of Polyakov loops wrap the Euclidean time circle in addition to the cycle in $\Sigma_g$.

Repeating the logic of the above analysis shows that the gravitational path integral in M-theory contains $N^g$ distinct supersymmetric black hole saddles. This multiplicity is the bulk manifestation of the $\big(\mathbb{Z}^{(1)}_N \big)^{g}$ electric one-form symmetry of the dual $\mathcal{N}=1$ SCFT, which in turn implies the presence of the term $g \log N$ in the effective action describing the black hole saddle of the gravitational index.

\subsubsection{Type IIA}
A similar analysis applies to type IIA string theory, both in its standard formulation (obtained by compactifying M-theory on a circle) and in massive IIA. In either case, the structure of the topological couplings inherited from the 10d theory determines the higher form symmetries realized in the holographic duals. 

AdS$_5$ solutions to massive IIA supergravity were found in \cite{Apruzzi:2015zna}. These feature non-trivial $H_3$ and $F_2$ fluxes, while $F_4$ is exact. 
The topological term for massive type IIA in the presence of non-trivial background fluxes requires some care, and it takes the form \cite{DeWolfe:2005uu,Cassani:2008rb}\footnote{The field strengths are given by:
$
\nonumber H = H_{\mathrm{fl}} + \diff B ,\,\,
F_2 = \diff C_1 + F^{\mathrm{fl}}_{2} + B\,F_0 ,\,\,
\nonumber F_4 = \diff C_3 - H \wedge C_1 + F^{\mathrm{fl}}_{4}
     + B \wedge F^{\mathrm{fl}}_{2}
     + \frac{1}{2} F_0 \,B \wedge B $.}:
\begin{align}\label{CSmIIA}
\nonumber \!\!\!\! \!\!\!\! \!\!\!\!  S_{\text{CS}} =& -\frac{1}{4}\int
C_3   H_{\text{fl}}   (\diff C_3 + 2F^{\text{fl}}_4)
+ B   (\diff C_3 + F^{\text{fl}}_4)  (\diff C_3 + F^{\text{fl}}_4)
+ B^2   F^{\text{fl}}_2   (\diff C_3 + F^{\text{fl}}_4)\\&
+ \frac{1}{3} B^3   F^{\text{fl}}_2   F^{\text{fl}}_2
+ \frac{1}{3} F_0\, B^3   (\diff C_3 + F^{\text{fl}}_4)
+ \frac{1}{4} F_0\, B^4  F^{\text{fl}}_2
+ \frac{1}{20} F_0^2 B^5 
\end{align}
Since the integral $\int H_3 = N_{H}$ is non-trivial and quantized, the first term in \eqref{CSmIIA} produces a coupling of the form $N_{H} \int_{M_7} C_3 \wedge \diff C_3$ in 7d. The discussion proceeds then precisely like for the M-theory case: expanding $C_3$ in a basis of cycles in the Riemann surface $\Sigma_g$, one obtains $c_2$ and $\tilde{c}_2$ two-forms, and the reduction of the 7d CS term includes the 5d CS term $N_{H} \int_{M_5} c_2\wedge \diff \tilde{c}_2$. This would describe a dual $\big(\mathbb{Z}_{N_H}^{(1)}\big)^g$ one-form symmetry.

Note that these solutions also have at least D6-brane sources and an $F_2$ flux in addition to a non-vanishing Romans mass $F_0$, however at least naively this does not seem to affect the relevant Chern--Simons term in the 5d effective theory. In particular, for the background fields configuration appearing in \cite{Apruzzi:2015zna}, the terms involving the Romans mass $F_0$ contain higher powers of the $B$-field which  vanish. The remaining terms involving $F_2^{\rm fl}$ do not contribute either, since the integral $\int B_2 \wedge F_2^{\rm fl}$ vanishes when evaluated over the internal space $S^2 \times \Sigma_g$.

\section{Outlook}\label{sec:conlcusions}

In this paper, we have explored the gravitational manifestation of the one-form symmetry underlying the degeneracy of black hole saddles in the superconformal index. Our analysis builds on and extends arguments originally given in~\cite{Aharony:1998qu}. We conclude by highlighting two directions for further investigation.

\paragraph{Supersymmetric order parameter for one-form symmetry?}
One may ask if there exist SCFT operators preserved by the supercharge which defines the superconformal index, and charged under the one-form symmetry. These operators could be used to distinguish the degenerate black hole saddles. Although inserting one such operator into the SCFT index at finite values of the chemical potentials would necessarily yield a vanishing result, its value on a given representative within the family of degenerate black hole saddles may be non-vanishing and computable.

The ordinary Polyakov loop, even after upgrading to a Wilson--Maldacena loop, does not preserve the needed supersymmetry. This is immediately seen by considering the Killing vector $\mathcal{K}$ constructed as a bilinear of the Killing spinor associated with the supercharge, and studying its action on the loop. Using the coordinates \eqref{eq:hatted_coords} introduced in the context of the black hole solution of section~\ref{sec3}, the Killing vector reads $\mathcal{K}= i\partial_{\hat t_E} + (1-\Omega_1)\partial_{\hat\phi}+ (1-\Omega_2)\partial_{\hat\psi}$, so it does not preserve a loop wrapping the Euclidean time circle and sitting at a point in $S^3$ \cite{Chen:2023lzq}. 
 One may spin up the loop so that it wraps the orbits of $\mathcal{K}$, with $\Omega_1,\Omega_2$ chosen such that the orbits close.
  Then one could try to evaluate the expectation value both at weak and strong 't Hooft coupling. The weak-coupling result for equal angular velocities may be given by the naive expression~\eqref{eq:Polyakov2}. Taking the large-$N$ limit of that expression in the black hole saddle $m=1$, $n=0$, 
we obtain $P/N = \,\frac{\sin(\pi  \tau)}{\pi  \tau}$, which simplifies to $P/N=1$ in the flat space limit $ \tau\to 0$.

Other supersymmetric operators that have been proposed as order parameters for the confinement/deconfinement transition are Gukov--Witten defect operators, realized in the bulk as partially wrapped D3-branes. These indeed take different expectation values in the   AdS$_5$ and in the black hole index saddles, distinguishing the two saddles~\cite{Chen:2023lzq,Cabo-Bizet:2023ejm}. However, such defects are uncharged under the one-form symmetry.

\paragraph{Gauging the one-form symmetry and magnetic theory.}
Our analysis can be repeated choosing different boundary conditions for the bulk two-forms. 
 In particular, imposing Dirichlet b.c.\ on the RR two-form and Neumann b.c.\ on the NSNS two-form would still realize the saddle degeneracy, with distinct saddles related through the action of a magnetic one-form symmetry. Indeed, after adding a boundary term and integrating by parts,
 and up to an overall sign, the term \eqref{csterm} takes the same form with $b_2$ and $c_2$ interchanged, so that the same mechanism described above produces now a degeneracy of saddles labelled by flat configurations of $c_2$ along the cigar-like two-cycle $\Sigma_2$. The corresponding one-form symmetry acts by shifting $\int_{\Sigma_2} c_2 \to \int_{\Sigma_2} c_2 +\frac{2 \pi k}{|G|}$, with $k=0,\ldots,|G|-1$. In the dual field theory, the operators charged under this symmetry are 't Hooft lines, as discussed in appendix \ref{app:bdry}, making the symmetry magnetic, consistently with the fact that interchanging $b_2$ and $c_2$ realizes electric-magnetic duality. 

The change in boundary conditions corresponds to gauging the electric one-form symmetry at the boundary. For instance, starting from $\mathcal{N}=4$ SYM with gauge group ${\rm SU}(N)$, gauging the one-form symmetry yields the quotient group ${\rm PSU}(N)={\rm SU}(N)/\mathbb{Z}_N$, where the elements of the center have been removed. The magnetic one-form symmetry is realized in the field theory through the nontrivial global structure of the gauge group. In particular, for ${\rm PSU}(N)$, it is encoded in the fundamental group
$\pi_1({\rm PSU}(N))=\mathbb{Z}_N$, which classifies ${\rm PSU}(N)$ bundles that do not lift to ${\rm SU}(N)$ bundles. The partition function of the ${\rm PSU}(N)$ gauge theory is related to that of the ${\rm SU}(N)$ gauge theory by 
$
    Z_{{\rm PSU}(N)} = N^{1-b_1}\, Z_{{\rm SU}(N)}\, ,
$
where $b_1$ is the first Betti number of the boundary manifold~\cite{Vafa:1994tf,Witten:1998wy}. For theories on
$S^1 \times S^3$, one has $b_1=1$, and $Z_{{\rm PSU}(N)} = \, Z_{{\rm SU}(N)}$, namely the indices are the same for the two theories. Thus, although the quotient by $\mathbb{Z}_N$ removes the electric one-form symmetry, the elements of $\pi_1({\rm PSU}(N))$ reinstate a $\mathbb{Z}_N$ global one-form symmetry, called magnetic. 

From the viewpoint of section \ref{sec:1form_sym_breaking}, the 3d EFT of the ${\rm PSU}(N)$ theory is obtained by gauging the one-form and zero-form symmetries descending from the electric one-form symmetry of the 4d ${\rm SU}(N)$ theory. Because the electric zero-form symmetry is gauged, the  configurations of the 3d scalars $u_i$ that labeled the $N$ distinct vacua in the ${\rm SU}(N)$ theory are now gauge-equivalent and should be identified. 

The 3d global symmetry is now given by magnetic $\mathbb{Z}_N^{(1)}$ and $\mathbb{Z}_N^{(0)}$ symmetries.
To obtain the magnetic zero-form symmetry in 3d, we gauge the electric one-form symmetry descending from the 4d theory by adding a dynamical two-form field $b_2$. Then, the gauging is implemented by adding a scalar $\tilde u$ and the topological term
$
S_{\rm top}=\frac{iN}{2\pi}\int_{S_3}\tilde u\, \diff b_2 
$.
Thus
\begin{align}
Z\sim
\int D b_2\, D\tilde u\,D \Phi\,
\exp\left[-S[\Phi]+\frac{iN}{2\pi}\int_{S^3}\tilde u\, \diff b_2
\right].
\end{align}
Since the 3-manifold is compact the $b_2$ field can have instantonic configurations such that $\int_{S^3} \diff b_2\in2 \pi \mathbb Z$. Summing over them implies $\tilde u=\frac{2 \pi k}{N}, \,\, k=0,\ldots,N-1$, and the partition function becomes
\begin{align}
Z\sim\sum_{k=0}^{N-1}\int  D b_2\,D \Phi\,\exp\left[-S[\Phi]+ik\int_{S^3}\, \diff b_2\right]\,,
\end{align}
making manifest the presence of $N$ configurations related through the shift $\tilde u \to \tilde u+\frac{2\pi k}{N}$.
This argument is similar to the one in the bulk discussion of section~\ref{sec:one_form_in_bulk}.

It would be interesting to study the intermediate cases with mixed boundary conditions, realizing the mixed electric-magnetic symmetry $\mathbb{Z}_P^{(1)} \times \mathbb{Z}_Q^{(1)}$ with $PQ=N$ \cite{Apruzzi:2021phx,Bergman:2022otk}, and the corresponding gravitational saddles.

\section*{Acknowledgments}

We would like to thank Andrea Boido for collaboration on the initial stages of this project, as well as Antonio Amariti, Alejandro Cabo-Bizet, Lorenzo Di Pietro, I\~naki Garc\'ia-Etxebarria, Ohad Mamroud, Pierluigi Niro, Ashoke Sen, Luigi Tizzano, Alberto Zaffaroni, and especially Fabio Apruzzi and Zohar Komargodski for very useful discussions and comments. We also thank Fabio Apruzzi for comments on the draft. This work was supported in part by MUR-PRIN contract 2022YZ5BA2 - Effective quantum gravity.

\appendix 
\section{The 5d model}\label{app:5dmodel}

\subsection{Consistent truncation of type IIB}\label{app:truncation}

We study a subsector of the consistent truncation of type IIB supergravity on five-dimensional Sasaki--Einstein structures given in \cite{Cassani:2010uw}. 
Referring to the notation used there, we set $ a_1^\Omega = a_2^\Omega = b^\Omega = c^\Omega = 0$ which implies $h_0^\Omega= g_0^\Omega = h_1^\Omega= g_1^\Omega = f_2^\Omega = f_3^\Omega =  0$. This solves the respective equations of motion and thus gives a consistent subtruncation.
 The remaining field strengths read
\begin{equation}
\begin{array}{lllllll}
h_3 \!\!&=&\!\! \diff b_2 - b_1\wedge \diff A\,,\qquad\qquad & g_3\; =&\!\! \diff c_2-c_1\wedge \diff A - C_0(\diff b_2 - b_1\wedge \diff A)\,, \\[3mm]
h_2 \!\!&=&\!\! \diff b_1\,, \;\;&g_2\; =&\!\! \diff c_1-C_0 \diff b_1\,,  \\[3mm]
h_1^J \!\!&=&\!\!  Db^J\,, &g_1^J\; =&\!\! D c^J - C_0 D b^J\,,
\end{array}
\end{equation}
\begin{eqnarray}
\nonumber &&f_1=Da +  \frac{1}{2}\big[\, b^JD c^J - \,\,b\leftrightarrow c\,\big],\qquad\quad
 f_2^J=\diff a_1^J  +\frac{1}{2}\left[\,b^J \diff c_1-b_1\wedge D c^J - \,\,b\leftrightarrow c\,\right],
 \end{eqnarray} 
where $ Db^J= \diff b^J -2b_1$, $D c^J=\diff c^J -2c_1$, $Da=\diff a - 2a_1^J-K A$.

The five-dimensional action reads:
\begin{equation}
	S \,=\, S_{\rm kin} + S_{\rm top} +  S_{\rm pot}\,,
	\label{5daction}
\end{equation}
with
\begin{eqnarray}
\nonumber
S_{\rm kin} \!\!\!\!&=&\! \displaystyle\frac{1}{2\kappa_{5}^2}\int_M \!\Big[ R  \star \!1 -\frac{28}{3}\diff U\wedge \star \diff U -\frac{8}{3}\diff U\wedge \star \diff V -\frac{4}{3}\diff V\wedge \star \diff V - \frac{1}{2}\rme^{\frac{8}{3}U+\frac{8}{3}V}\diff A\wedge \star \diff A\\ [4mm] 
\nonumber -\!\!\!&\displaystyle\frac{1}{2}&\!\!\!\!\!  \diff \phi\wedge \star  \diff \phi -\frac{1}{2} \rme^{2\phi}\diff C_0 \wedge \star  \diff  C_0  -  \rme^{-\frac{4}{3}U-\frac{4}{3}V}  f_2^J\wedge \star  f_2^J - 2\rme^{-8U} f_1\wedge \star  f_1\\ [4mm]
\nonumber -\!\!\!&\displaystyle\frac{1}{2}&\!\!\!\! \rme^{-\phi}   \left(\rme^{\frac{16}{3}U+\frac{4}{3}V}   h_3\wedge \star h_3 + \rme^{\frac{8}{3}U-\frac{4}{3}V} h_2\wedge \star h_2+2\rme^{-4U}h_1^J\wedge \star h_1^J \right) \\[4mm]
\label{Skin}
-\!\!\!&\displaystyle\frac{1}{2}&\!\!\!\! \left.\rme^{\phi}   \left(\rme^{\frac{16}{3}U+\frac{4}{3}V}\,g_3\wedge \star g_3+\rme^{\frac{8}{3}U-\frac{4}{3}V}g_2\wedge \star g_2+2\rme^{-4U}g_1^J\wedge \star g_1^J  \right)\right],
\end{eqnarray}
\begin{eqnarray} 
\nonumber S_{\rm top} \!\!&=& \displaystyle\frac{1}{2\kappa_{5}^2}\int_M \left\{  A\wedge \diff a_1^J \wedge \diff a_1^J  \right. \label{SCS}\\ [4mm]
	 \!\!\!\!\!\!\!\!&& \displaystyle-\,\frac{1}{2}\big(\,\diff a_1^J + f_2^J\,\big)\wedge\big[b_2\wedge Dc^J+b^J\left(\diff c_2-c_1\wedge \diff A\right)- \,\,b\leftrightarrow c\,\,\big]\\[4mm]
\nonumber \!\!\!\!\!\!\!\!&&  \displaystyle+\,\frac{1}{2}\left(Da+f_1\right)\wedge\left[b_2\wedge \diff c_1-b_1\wedge\left(\diff c_2-c_1\wedge \diff A\right)-\,\,b\leftrightarrow c\,\,\right]\\ [4mm]
\nonumber	 \!\!\!\!\!\!\!\!&& -\,K\left[b_2\wedge \left(\diff c_2-c_1\wedge \diff A\right)-\,\,b\leftrightarrow c\,\,\right]\bigg\}\,,
\end{eqnarray}
\begin{equation}
\begin{array}{rcl}
S_{\rm pot} \:=\; \displaystyle\frac{1}{2\kappa_{5}^2}\int_M \big(-2\mathcal V\big) \star 1 \!\!\!&=&\!\!\! \displaystyle\frac{1}{2\kappa_{5}^2}\int_M \left[ 24\, \rme^{-\frac{14}{3}U-\frac{2}{3}V} - 4\, \rme^{-\frac{20}{3}U + \frac{4}{3}V} -2\, \rme^{-\frac{32}{3}U -\frac{8}{3}V} K^2 \right]\star 1\,.
\end{array}
\label{Spot}
\end{equation}
The 5-dimensional gravitational coupling constant is obtained from the ten-dimensional one $\kappa_5^2 \equiv \kappa_{10}^2/{\rm Vol}$, where ${\rm Vol}$ is the standard volume of the internal Sasaki--Einstein manifold. 
The gauge transformations of these 5d fields, including the action of the one-form gauge parameters for the two-form potentials, were described in appendix~C of~\cite{Cassani:2010na}.

\subsection{Boundary conditions}\label{app:bdry}
Let us discuss the variational principle associated with the model we are considering. This requires
fixing boundary conditions at asymptotic infinity such that the action (given by the bulk
action plus possible boundary terms that may be added) is extremized upon imposing the
bulk equations of motion.

In particular, we wish to determine the boundary conditions to be imposed on $b_2$ and
$c_2$. We will rename the real scalars $b_J \rightarrow b_0$, $c_J \rightarrow c_0$. We only consider the
terms in the action above which contain either $b_2$ or $c_2$.
These terms are
\begin{align}
&S_{\text{kin}} = \frac{\rme^{\frac{16}{3} \hat{U}+\frac{4}{3} \hat{V}}}{2\kappa_5^2} \int \Big[
-\frac{1}{2} \rme^{-\phi} \big( (\mathrm{d}b_2 - b_1 \wedge \mathrm{d}A) \wedge \star (\mathrm{d}b_2 - b_1 \wedge \mathrm{d}A) \big)
\nonumber\\
&
-\frac{1}{2} \rme^{\phi} \big( (\mathrm{d}c_2 - c_1 \wedge \mathrm{d}A - C_0(\mathrm{d}b_2 - b_1 \wedge \mathrm{d}A)) \wedge
\star (\mathrm{d}c_2 - c_1 \wedge \mathrm{d}A - C_0(\mathrm{d}b_2 - b_1 \wedge \mathrm{d}A)) \big)
\Big],
\\[4mm]
&S_{\text{top}} = \frac{1}{2\kappa_5^2} \int \Big\{
-\frac{1}{2}(\mathrm{d}a_{1}^J + f_{2}^J)
\wedge \big[
b_2 \wedge D c_0 + b_0(\mathrm{d}c_2 - c_1 \wedge \mathrm{d}A) - b \leftrightarrow c
\big]
\nonumber\\
&
+\frac{1}{2}(D a + f_1) \wedge
\big[b_2 \wedge \mathrm{d}c_1 - b_1 \wedge (\mathrm{d}c_2 - c_1 \wedge \mathrm{d}A) - b \leftrightarrow c \big]
- K \big[b_2 \wedge (\mathrm{d}c_2 - c_1 \wedge \mathrm{d}A) - b \leftrightarrow c \big]
\Big\}.
\end{align}

We would like to remove the terms ``$-b \leftrightarrow c$'' from $S_{\text{top}}$ above, both the ones appearing
explicitly and those appearing implicitly in the definition of $f_1$ and $f_{2}^J$. In order to do so,
we redefine $a$ and $a_{1}^J$ as
\begin{equation}
D a \,\to\, D a + \tfrac{1}{2}(c_0 D b_0 + b_0 D c_0), \qquad\quad
a_1^J \,\to\, a_1^J + \tfrac{1}{2}(b_0 c_1 + c_0 b_1),
\label{eq:a_redef}
\end{equation}
(the first follows from $a \to a + \tfrac{1}{2} b_0 c_0$), which also gives the following new expressions:
\begin{equation}
f_1 \,=\, D a + b_0 D c_0\,, \qquad \quad f_2^J \,=\, \diff a_1^J + b_0 \mathrm{d}c_1 - b_1 \wedge D c_0\,.
\label{eq:new_f}
\end{equation}
Here, $D a$, $D c_0$ and $D b_0$ have the same expressions as before the redefinition.
Implementing integrations by parts and taking into account several recombinations, the
topological term now reads
\begin{align}
S_{\text{top}} &= \frac{1}{2\kappa_5^2} \int \Big\{
-(\mathrm{d}a_{1}^J + f_{2}^J)
\wedge \big[
b_2 \wedge D c_0 + b_0(\mathrm{d}c_2 - c_1 \wedge \mathrm{d}A)
\big]
\nonumber\\
&\qquad
+ (D a + f_1) \wedge
\big[
b_2 \wedge \mathrm{d}c_1 - b_1 \wedge (\mathrm{d}c_2 - c_1 \wedge \mathrm{d}A)
\big]
- 2K \big[ b_2 \wedge (\mathrm{d}c_2 - c_1 \wedge \mathrm{d}A) \big]
\Big\}.
\end{align}
We assume our model is the bulk action above, with no boundary terms being added.

We now consider the first order variation of the action. We call $b_2', c_2'$ the variations of $b_2, c_2$ around any given configuration. The boundary terms generated by partial
integration when considering the variation of the action are:
\begin{align}
\nonumber S_{\text{bdry}} = \frac{1}{2\kappa_5^2} \int_{\partial M} &\Big[
 - \rme^{\frac{16}{3} \hat{U}+\frac{4}{3} \hat{V}}\rme^{-\phi} b_2' \wedge \star (\mathrm{d}b_2 - b_1 \wedge \mathrm{d}A)\\
&-\rme^{\frac{16}{3} \hat{U}+\frac{4}{3} \hat{V}} \rme^{\phi}(c_2' - C_0 b_2') \wedge
\star\big( (\mathrm{d}c_2 - c_1 \wedge \mathrm{d}A) - C_0(\mathrm{d}b_2 - b_1 \wedge \mathrm{d}A) \big)
\nonumber\\
& - \big[
(2\,\mathrm{d}a_1^J + b_0 \mathrm{d}c_1) b_0
- 2 b_0 b_1 \wedge D c_0
+ 2 D a \wedge b_1
+ 2K\, b_2
\big] \wedge c_2'
\Big].
\label{eq:Sbdry}
\end{align}
We fix boundary conditions for $b_2$ such that
\begin{equation}
b_2'|_{\partial M} = 0\,,
\label{eq:bc_b2}
\end{equation}
and in addition choose $b_2|_{\partial M}$ such that the term proportional to $c_2'$ vanishes at the boundary:
\begin{align}\label{current}
\nonumber \star j_c \equiv \star J_c|_{\partial M}=
&\frac{1}{2 \kappa_5^2}\Big[
\rme^{\frac{16}{3} \hat{U}+\frac{4}{3} \hat{V}} \rme^{\phi} \star \big( (\mathrm{d}c_2 - c_1 \wedge \mathrm{d}A) - C_0 (\mathrm{d}b_2 - b_1 \wedge \mathrm{d}A) \big)\\
&+ (2\,\mathrm{d}a_{1}^J + b_0 \mathrm{d}c_1)b_0
- 2 b_0 b_1 \wedge D c_0
+ 2 D a \wedge b_1
+ 2K\, b_2
\Big]_{\partial M} = 0\,.
\end{align}

Here, $j_c$ is the boundary conserved current associated with gauge transformations
$c_2 \to c_2 + \mathrm{d}\gamma_1$ (we will instead call $j_b$ the current associated with
$b_2 \to b_2 + \mathrm{d}\lambda_1$).

This choice is compatible with the $c_2$ equation of motion
evaluated at the boundary. Indeed the $c_2$ equation of motion is just the exterior derivative
of the expression above (evaluated in the bulk), hence it is solved by the expression
above up to an arbitrary closed two-form. Our boundary condition amounts to
demanding that this arbitrary two-form vanishes at the boundary.

By demanding $\star j_c = 0$ we are demanding vanishing of the charge associated with such
current. From a dual field theory point of view, the $c_2$ gauge transformations are those of
a magnetic one-form symmetry, hence we are forbidding a magnetic one-form charge.

We also note that if we neglect all two-derivative terms in the five-dimensional action
and only keep the one-derivative Chern–Simons term, the condition~\eqref{current} boils down to
$b_2|_{\partial M} = 0$, which is the one appearing in \cite{Hofman:2017vwr}.
\paragraph{Symmetry generators and charged operators.} From varying the action, one finds:
\begin{align}\label{currentcons}
    \diff \star J_b=0\,, \quad\qquad\diff \star J_c=0\,,
\end{align}
which can be interpreted as conservation equations for the bulk currents $J_b$ and $J_c$, where
\begin{align}
\nonumber \star J_b\,=\frac{1}{2 \kappa_5^2} \Big[\,&\rme^{\frac{16}{3} \hat{U}+\frac{4}{3} \hat{V}}\Big( \left(\rme^{-\phi}+\rme^\phi C_0^2\right)\,\star\,(\diff b_2-b_1 \wedge \diff A)-C_0 \,\rme^\phi \star(\diff c_2-c_1 \wedge \diff A)\Big)\\
&-2 \,\diff a_1^J\, c_0+2 a \,\diff c_1-2K \,c_2\Big].
\end{align}
These currents are not gauge invariant due to the presence of bare gauge fields in their expressions. Despite this, one can construct gauge-invariant operators by taking the exponential of their fluxes over closed surfaces:
\begin{equation}\label{operator}
    Q_s(\Sigma_2)\,=\,\exp\!\left(\frac{2 \pi i\, s}{N}\int_{\Sigma_2}\star J_b\right), \qquad  \tilde Q_r(\tilde\Sigma_2)\,=\,\exp\!\left(\frac{2 \pi i\, r}{N}\int_{\tilde\Sigma_2}\star J_c\right)\,,
\end{equation}
where $\Sigma_2, \tilde \Sigma_2$ are closed two-dimensional surfaces in the bulk. These operators are well-defined for $r,s\in\mathbb Z$ and topological, due to \eqref{currentcons}. They are symmetry defect operators (SDO) which implement the symmetry by acting respectively on the charged operators:
\begin{align}\label{wilson}
    &\mathcal{W}_n({\Omega_2})\,=\,\rme^{in\int_{\Omega_2} b_2}, \qquad  \tilde{\mathcal{W}}_m(\tilde\Omega_2)\,=\,\rme^{im\int_{\tilde{\Omega}_2}  c_2} \qquad\qquad m,n \in \mathbb{Z}\,,
\end{align}
where $\Omega_2,\tilde{\Omega}_2$ are surfaces ending on the boundary, namely $\partial \Omega_2, \partial \tilde{\Omega}_2 \subset \partial M$. 

We now show that these operators carry the correct charges under the SDOs. To this end, consider inserting a charged operator $\mathcal{W}_n(\Omega_2)$ in the path integral. In its presence, the equation of motion for $b_2$ is modified to:
\begin{align}\label{wi}
     \diff \star J_b\,=\, n \,\delta^{(3)}(\Omega_2)\,,
\end{align}
where $\delta^{(3)}(\Omega_2)$ is a delta-function three-form localized on $\Omega_2$, normalized such that it integrates to one over any three-dimensional manifold that intersects $\Omega_2$ once.
 
Then, we consider a charged operator of charge $n$, supported on a two-dimensional surface $\Omega_2$ and an SDO supported on a two-dimensional surface $\Sigma_2$ which links with $\Omega_2$. We can continuously deform $\Sigma_2 \rightarrow \Sigma_2'$ so that $\Sigma_2'$ does not link with $\Omega_2$. Introducing a three-dimensional manifold $\hat{\Sigma}_3$ such that $\partial \hat{\Sigma}_3=\Sigma_2-\Sigma_2'$, and using \eqref{wi} we obtain:

\begin{align}\label{actionsymmetry1}
\nonumber \braket{Q_s(\Sigma_2)\mathcal{W}_n(\Omega_2)}&=\,\int Db_2 Dc_2  \,\rme^{iS}\rme^{\frac{2\pi i s}{N}\int_{\Sigma_2}\star J_b} \rme^{in \int_{\Omega_2}b_2}\\
\nonumber    &=\,\int Db_2 Dc_2  \,\rme^{iS} \rme^{\frac{2 \pi i s}{N}\int_{\hat{\Sigma}_3}\diff \star J_b}\rme^{in \int_{\Omega_2}b_2}\rme^{\frac{2 \pi i s}{N}\int_{\Sigma_2'}\star J_b}\\
\nonumber    &=\,\int Db_2 Dc_2  \,\rme^{iS} \rme^{\frac{2 \pi i s}{N}\int_{\hat{\Sigma}_3} n \,\delta^{(3)}(\Omega_2)}\rme^{in \int_{\Omega_2}b_2}\rme^{\frac{2 \pi i s}{N}\int_{\Sigma_2'}\star J_b}
    \\
    &=\,\exp{\bigg(\frac{2\pi ins}{N} \,\text{Lk}(\Sigma_2,\Omega_2)\bigg)}\braket{\mathcal{W}_n(\Omega_2)Q_s(\Sigma_2')}\,,
\end{align} 
where we used that: 
\begin{align}
    \int_{\hat{\Sigma}_3}\delta^{(3)}(\Omega_2)\,=\,\text{Lk}(\Sigma_2,\Omega_2)\,.
\end{align}
Analogously, one can show:
\begin{align}\label{actionsymmetry2}   
\braket{\tilde Q_r(\tilde\Sigma_2)\tilde{\mathcal{W}}_m(\tilde\Omega_2)}\,=\,\exp{\bigg(\frac{2\pi imr}{N} \,\text{Lk}(\tilde\Sigma_2,\tilde\Omega_2)\bigg)}\braket{\tilde{\mathcal{W}}_m(\tilde\Omega_2)\tilde Q_r(\tilde\Sigma_2')}\,.
\end{align}
We can verify that the symmetry of the theory is actually $\mathbb{Z}_N^{(1)}$ by observing that the action of an SDO on a charged operator of charge $n$ is equivalent to the action on a charged operator of charge $n+N$:
\begin{align} \label{eq:commrelQW}
\nonumber    \braket{Q_s(\Sigma_2)\mathcal{W}_{n+N}(\Omega_2)}\,&=\,\exp{\bigg(\frac{2\pi i(n+N)s}{N} \,\text{Lk}(\Sigma_2,\Omega_2)\bigg)}\braket{\mathcal{W}_n(\Omega_2)Q_s(\Sigma_2')}\\
\nonumber    &=\,\exp{\bigg(\frac{2\pi ins}{N} \,\text{Lk}(\Sigma_2,\Omega_2)\bigg)}\braket{\mathcal{W}_n(\Omega_2)Q_s(\Sigma_2')}\\[1mm]
    &=\,\braket{Q_s(\Sigma_2)\mathcal{W}_n(\Omega_2)}\,,
\end{align}
implying, thus, that an operator of charge $N$ transforms trivially. The same statement holds for the operators $\tilde{\mathcal{W}}$ and $\tilde{Q}$. 

\paragraph{Boundary conditions and gauge group of the boundary theory.} By imposing the asymptotic behaviour specified in \eqref{eq:bc_b2} and \eqref{current}, we enforce Dirichlet boundary conditions on $b_2$ and Neumann-like boundary conditions on $c_2$. As a result, $c_2$ is a dynamical field  on the boundary. The bulk field $b_2$ couples electrically to fundamental strings, whose world-sheets can end on the boundary as line operators—interpreted as Wilson lines—while $c_2$ couples to D1-branes, which can end on the boundary producing objects interpreted as 't Hooft lines, according to electric-magnetic duality.

After imposing \eqref{current}, the only surviving topological operators on the boundary are $Q_s(\Sigma_2)$, which act on $\mathcal{W}_n(\Omega_2)$ via \eqref{actionsymmetry1}. This induces an electric $\mathbb{Z}_N$ one-form symmetry acting on the Wilson lines. In the boundary theory this means that the center is not gauged, and the gauge group is ${\rm SU}(N)$.

For Neumann-like boundary conditions on $b_2$ and Dirichlet on $c_2$, instead, the surviving operators on the boundary are $\tilde Q_r(\tilde \Sigma_2)$, generating again a $\mathbb{Z}_N$ one-form symmetry which acts on the $\tilde{\mathcal{W}}_m(\tilde\Omega_2)$. In this case, being the charged objects interpreted as 't Hooft lines on the boundary, the global structure of the boundary gauge group is correspondingly fixed to be ${\rm PSU}(N)\equiv {\rm SU}(N)/\mathbb Z_N$.

From the holographic viewpoint, the choice of boundary conditions for $(b_2,c_2)$ determines which of the two discrete $\mathbb{Z}_N^{(1)}$ symmetries is gauged and which remains global:
\begin{itemize}
    \item \textbf{Dirichlet on $b_2$}: the boundary value of $b_2$ is fixed and acts as a background field for the electric $\mathbb{Z}_N^{(1)}$ global symmetry;
    \item \textbf{Dirichlet on $c_2$}: the boundary value of $c_2$ is fixed and acts as a background field for the magnetic $\mathbb{Z}_N^{(1)}$ global symmetry;
    \item \textbf{Mixed b.c.}: $Q\, b_2$ and $P\,c_2$ are fixed at the boundary and act as background fields for a $\mathbb{Z}_P^{(1)} \times \mathbb{Z}_Q^{(1)}$ symmetry, where $PQ=N$.
\end{itemize}
Near the boundary, the only relevant term is the topological term \eqref{eq:bdc_term},
implying that $b_2$ and $c_2$ are canonically conjugate variables. As a result, one cannot impose boundary conditions that gauge both $\mathbb{Z}_{N}^{(1)}$ symmetries.

\subsection{Fluctuations} \label{app:fluct}

In the main text, we are interested in studying the fluctuations of the type IIB supergravity fields around the Euclidean black brane solution \eqref{ARmetric}. The latter is originally constructed in five dimensions. Although it would be straightforward to uplift it to ten dimensions, we choose a slightly different approach: we place ourselves in the five-dimensional consistent truncation described above; this means we restrict to a subset of all possible fluctuations of the type IIB fields. 
Starting from the action reported in section \ref{app:truncation}, we first remove the ``$-b \leftrightarrow c$'' terms by redefining $a,a_1^J$, as explained in \ref{app:bdry}. Then, since we are interested in studying the dynamics of $b_2$ and $c_2$ we keep only their kinetic term and the part of $S_{\rm top}$ in which they appear.
Altogether, the action reads:\footnote{Notice that the fields of the 10d parent theory $B_2$ and $C_2$ used here are related to the ones appearing in the usual type IIB SUGRA action (which we call $B^0_2,C_2^0$) as $B_2 = \frac{B^0_2}{2\pi}$, $C_2 = \frac{C^0_2}{2\pi}$, so that the Chern--Simons term $\int b_2 \wedge \diff c_2$ is multiplied by the usual prefactor $\frac{N}{2\pi}$. This explains the overall $(2\pi)^2$ which was not present in \cite{Cassani:2010uw}.}
\begin{align}\label{action}
\nonumber  S&=\frac{(2\pi)^2}{2 \kappa_5^2} \int -\frac{1}{2} \rme^{-\phi} \rme^{\frac{16}{3} U+\frac{4}{3} V}(\diff b_2-b_1\wedge \diff A)\wedge \star (\diff b_2-b_1 \wedge \diff A)\\[4mm]
\nonumber    &-\frac{1}{2} \rme^{\frac{16}{3} U+\frac{4}{3}V+\phi} \left[(\diff c_2-c_1 \wedge \diff A)-C_0(\diff b_2-b_1 \wedge \diff A)\right] \wedge \star \left[(\diff c_2-c_1 \wedge \diff A)-C_0(\diff b_2-b_1 \wedge \diff A)\right]\\[4mm]
\nonumber    &-\left[2 \diff a_1^J+b_0 \diff c_1-b_1 \wedge (\diff c_0-2c_1)\right]\wedge \left[b_2 \wedge (\diff c_0-2c_1)+b_0(\diff c_2-c_1 \wedge \diff A)\right]\\[4mm]
\nonumber &+\left[2 \diff a-4a_1^J-2K A+b_0 (\diff c_0-2c_1)\right]\wedge \left[b_2 \wedge \diff c_1-b_1\wedge(\diff c_2-c_1 \wedge \diff A)\right]\\[4mm]
    &-2K \left[b_2 \wedge (\diff c_2-c_1 \wedge \diff A)\right].
\end{align}
The first two terms describe the gauge-invariant kinetic terms of $b_2$ and $c_2$, while the last term, proportional to $K$, contains the topological coupling $b_2\wedge \diff c_2$. In addition, since the background field strength $\diff A$ is non-vanishing, there appear extra interaction terms mixing the two-forms with $A$ and with the one-form fields $b_1$, $c_1$ and the scalars $b_0,c_0$. 

We adopt the field expansion introduced earlier.
Any field $\Phi$ of the 5d theory is written as
$\Phi = \vev{\Phi} + \Phi'$, where $\vev{\Phi}$ is the classical
background solving the equations of motion and $\Phi'$ denotes
the fluctuation around it.

The configuration \eqref{ARmetric} is a solution to minimal gauged supergravity. This requires $\vev{C_0}, \vev{\phi}, \vev{a},\vev{b}^J,\vev{c}^J$ constant, $\vev{U} = \vev{V} =0$, $\vev{b}_1=\vev{c}_1=0$, $\vev{a}_1^J = -\vev{A}$. Importantly, we take $\vev{h}_3=\vev{g}_3=0$ but we allow for non-vanishing, flat $\vev{b}_2,\vev{c}_2$.
That is, we take $b_2 = \vev{b}_2 + b_2'$, $c_2 = \vev{c}_2 + c_2'$, with $\diff \vev{b}_2 =\diff \vev{c}_2= 0$.
We plug this into the action and keep only the terms that are at most quadratic in the fluctuations. The zeroth-order term is simply the action evaluated in the background (this also contains the holographic counterterms), while the first-order term vanishes upon imposing the boundary conditions as in section~\ref{app:bdry}.
In order to analyze the vacuum structure and its possible degeneracy, we focus on the fluctuations $c_2'(\mathbf{x})$ and $b_2'(\mathbf{x})$ whose two indices are both along the non-compact directions $\mathbb{R}^3$ and which depend only on the $\mathbb{R}^3$ coordinates $\mathbf{x}$. This restriction should be understood as a projection onto the zero-mode sector of the two-form fields. Since we are interested in global properties of the vacuum rather than in propagating degrees of freedom, it is sufficient to retain modes that are constant on $D_2$. In contrast, fluctuations such as $b_2'$ with both legs along the internal $D_2$ do not couple to $\diff \hat{A}= \frac{3\varrho^3 g^2}{r'^3 } \diff r' \wedge \diff \zeta$ and therefore do not probe the charge carried by the configuration.
We obtain:
\begin{eqnarray}
\nonumber
 \delta S&&\!\!\!\!\!\!\!\!\!+\delta^2 S =\frac{(2\pi)^2}{2 \kappa_5^2} \int -\frac{1}{2} \rme^{-\hat{\phi}} \rme^{\frac{16}{3} \hat{U}+\frac{4}{3} \hat{V}}(\diff b_2'-b_1'\wedge \diff \hat{A})\wedge \star (\diff b_2'-b_1' \wedge \diff \hat{A})\\[4mm]
\nonumber &&\!\!\!\!\!\!\!\!\!\!\!\!\!\!-\frac{1}{2} \rme^{\hat{\phi}}\rme^{\frac{16}{3} \hat{U}+\frac{4}{3}\hat{V}} \left[(\diff c_2'-c_1' \wedge \diff \hat{A})-\hat{C}_0(\diff b_2'-b_1' \wedge \diff \hat{A})\right] \wedge \star \left[(\diff c_2'-c_1' \wedge \diff \hat{A})-\hat{C}_0(\diff b_2'-b_1' \wedge \diff \hat{A})\right]\\[4mm]
\nonumber   &&\!\!\!\!\!\!\!\!\!\!\!\!\!\! +2\, \diff \hat{A}\wedge \left[(\hat{b}_2+b_2') \wedge \diff c_0'+(\hat{b}_0+b_0')\,\diff c_2'\right] -\hat{b}_0 \, \diff c_1'\wedge \left[\hat{b}_2 \wedge (\diff c_0'-2c_1')+\hat{b}_0(\diff c_2'-c_1' \wedge \diff \hat{A})\right]\\[4mm]  &&\!\!\!\!\!\!\!\!\!\!\!\!\!\!+\hat{b}_0 \,(\diff c_0'-2c_1')\wedge \left(\hat{b}_2 \wedge \diff c_1'\right)-2K\, \hat{b}_2 \wedge \diff c_2'\,,
\end{eqnarray}
where the metric entering in the Hodge star in the first line is the background one. 
It is convenient to set $\vev{b}_0=\vev{c}_0=0$, although not strictly required. Then we obtain
\begin{eqnarray}
\nonumber
 \delta S&&\!\!\!\!\!\!\!\!\!\!+\delta^2 S =\frac{(2\pi)^2}{2 \kappa_5^2} \int -\frac{1}{2} \rme^{-\hat{\phi}} \rme^{\frac{16}{3} \hat{U}+\frac{4}{3} \hat{V}}(\diff b_2'-b_1'\wedge \diff \hat{A})\wedge \star (\diff b_2'-b_1' \wedge \diff \hat{A})\\[4mm]
\nonumber &&\!\!\!\!\!\!\!\!\!\!\!\!\!\!\!\!-\frac{1}{2} \rme^{\hat{\phi}}\rme^{\frac{16}{3} \hat{U}+\frac{4}{3}\hat{V}} \!\left[(\diff c_2'-c_1' \wedge \diff \hat{A})-\hat{C}_0(\diff b_2'-b_1' \wedge \diff \hat{A})\right] \wedge \star \left[(\diff c_2'-c_1' \wedge \diff \hat{A})-\hat{C}_0(\diff b_2'-b_1' \wedge \diff \hat{A})\right]\\[4mm]
 &&\!\!\!\!\!\!\!\!\!\!\!\!\!\!\!\! +\,2 \, \diff \hat{A}\wedge \left[(\hat{b}_2+b_2') \wedge \diff c_0'+b_0' \,\diff c_2'\right]-2K\, \hat{b}_2 \wedge \diff c_2'\,.
\end{eqnarray}
The first two lines, together with the last term in the third line, give the terms relevant for the argument of \cite{Aharony:1998qu}. The presence of charge in the black hole solution \eqref{CCLP}, instead, gives rise to additional couplings involving the background gauge field $\vev{A}$.

\setstretch{1.1}

\bibliographystyle{JHEP}
\bibliography{dual_of_logN.bib}

\end{document}